\documentclass[10pt,letterpaper,journal, twocolumn, final]{IEEEtran}
\usepackage{amsmath, amssymb,mathtools,physics}
\usepackage{graphicx}
\usepackage[inline]{enumitem}
\usepackage{pgfplots}
\pgfplotsset{compat=1.15}
\usepackage{mathrsfs}
\usepackage{multirow}
\usepackage{enumitem}
\usepackage[]{booktabs}

\usepackage{subcaption}
\DeclareMathOperator*{\argmin}{arg\,min}

\author{\textbf{Nikhil ~Sharma},~ \and\textbf{Shovan ~Bhaumik},~  \and \textbf{Ratnasingham ~Tharmarasa},~ \and \textbf{Thia ~Kirubarajan}}
\title{Trigonometric Moments of a Generalized von Mises Distribution in 2-D Range-Only Tracking.}

\graphicspath{{images/sim2}{images/compare_quadrature/}{images/determSampling/}}
\begin{document}
	\maketitle
	\begin{abstract}
		A 2D range-only tracking scenario is non-trivial due to two main reasons. First, when the states to be estimated are in Cartesian coordinates, the uncertainty region is multi-modal. The second reason is that the probability density function of azimuth conditioned on range takes the form of a generalized von Mises distribution, which is hard to tackle. Even in the case of implementing a uni-modal Kalman filter, one needs expectations of trigonometric functions of conditional bearing density, which are not available in the current literature. We prove that the trigonometric moments (circular moments) of the azimuth density conditioned on range can be computed as an infinite series, which can be sufficiently approximated by relatively few terms in summation. The solution can also be generalized to any order of the moments.
		
		This important result can provide an accurate depiction of the conditional azimuth density in 2D range-only tracking geometries. We also present a simple optimization problem that results in deterministic samples of conditional azimuth density from the knowledge of its circular moments leading to an accurate filtering solution. The results are shown in a two-dimensional simulation, where the range-only sensor platform maneuvers to make the system observable. The results prove that the method is feasible in such applications.
		 
	\end{abstract}

	\section{Introduction}

	\IEEEPARstart{T}{racking} with both azimuth and range measurements is considered as a benchmark problem in target tracking scenarios. Due to the evolution of low cost sensing and embedded technologies, tracking with range-only measurements has become a recent interest since the last decade. Also, due to advances in radio frequency technology, measuring range between beacons and transponder is inexpensive and efficient. This has lead to a tremendous use of range-only measurements in wireless sensor networks (WSN) \cite{kantor2002preliminary}. In this work, we consider a single-sensor 2D range-only target motion analysis, which is considered impractical in long-range due to three primary reasons. 
	\begin{enumerate}[label*=(\arabic*)]
		\item Gaussian assumption is invalid for most scenarios since the uncertainty region resembles a concave shape. For a sensor with a large field-of-view (FOV) and low standard deviation, this cannot be sufficiently approximated with an ellipsoid \cite{tian2014consistency, clark2010gaussian, clark2009gaussian}. 
		\item The density of azimuth conditioned on range is extremely hard to simulate due to complicated expression and non-trivial integrals. 
		\item While the observability conditions have been established for observer motion with several legs, there is no generalized solution for the observability of range-only target motion analysis \cite{pillon2016observability, song1999observability}.		
	\end{enumerate}
	
	The first problem has been solved in various cases by either parameterization of azimuth \cite{ristic2002target} or using a multi-modal density instead of Gaussian. A Gaussian mixture approach was used in \cite{clark2010gaussian} and \cite{clark2009gaussian} wherein, both the prior density and the conditional azimuth density were assumed to be Gaussian mixtures. The authors used EM algorithm for Gaussian mixture approximation from the knowledge of azimuth density, but the azimuth density itself was sampled uniformly.
	
	Due to the observability issues, the range-only tracking is seldom used except in WSNs wherein measurements from multiple nodes are fused together to achieve observability. As mentioned before, unlike bearings-only target motion analysis, observability in range-only scenarios do not possess a general solution for observer motion. Pillion et al. in \cite{pillon2016observability} analyzed leg-by-leg observer trajectory and found that a multi-leg scenario might be observable if the range-rate is constant. The analysis was performed after assuming a nearly constant velocity (NCV) model for the target. 
	
	In this paper, we focus on the first two problems which are based on the multi-modal shape of the density of azimuth conditioned on range measurements, in two-dimensional case. Specifically, we derive a generalized solution for the conditional integral of the form,
	 { \begin{subequations}\label{eq_mainIntegralProbs}
		 \begin{align}
		 	\mathbb{E}\left[\cos(m\theta)|r\right] &= \int_{0}^{2\pi} \cos(m\theta) p(\theta|r) d\theta, \quad m = 1,2,\dots \label{eq_int_cosm}\\
		 	\mathbb{E}\left[\sin(m\theta)|r\right] &= \int_{0}^{2\pi} \sin(m\theta) p(\theta|r) d\theta, \quad m = 1,2,\dots \label{eq_int_sinm}
		 \end{align}
 	\end{subequations}}
	 Such integrals are better known as the trigonometric moments or circular moments in literature. Here, $\theta$ is the azimuth, $r$ is the range,  $m$ is a positive integer and $p(.)$ represents the probability density function. We show that the conditional density $p(\theta|r)$ resembles a generalized form of the von Mises distribution \cite{gatto2007generalized}. As per the author's knowledge, trigonometric moments of such a distribution do not exist in the current literature, and hence are a powerful tool to infer the shape of conditional azimuth density in a range-only tracking scenario. 
	 
	 The authors present a generalized solution such that trigonometric moment of any order can be calculated with sufficient accuracy. Based on this, the deterministic samples of the conditional azimuth density can be determined using a simple optimization-based approach. The results are directly applied to a 2D range-only target-tracking problem.
	 
	 The rest of the paper is organized as follows : In Section \ref{sec_probForm} we provide a general overview of the range-only tracking problem along with a modified  measurement model. Generalized update equations for range-only tracking are presented in Section \ref{sec_updateEqn}. We present an unimodal Kalman filter type recursion for range-only scenarios in this section. The true range density and the conditional azimuth densities are derived in Section \ref{sec_condAzimuth_density}. We solve the integrals in Eqn. \eqref{eq_mainIntegralProbs} in Section \ref{sec_expOfCond_density} and subsequently apply it to find deterministic samples of the conditional density in Section \ref{sec_detSample}.
	 The derivations are tested against numerical quadrature and a 2D simulation in Section \ref{sec_sim}. Finally, the article is concluded in Section \ref{sec_concl}, followed by an Appendix which contains proof of the formulations presented.   
	 
	 \section{Problem Formulation} \label{sec_probForm}
		Consider a linear Gaussian system which can be represented in discrete time state-space formulation as,
	\begin{align}
		\mathbf{x}^r_k = \mathbf{F}_k\mathbf{x}^r_{k-1} + \mathbf{F}_k\mathbf{x}^o_{k-1} - \mathbf{x}^o_{k} +  \mathbf{w}_k \label{eq_stateEq},
	\end{align}
	where $\mathbf{x}^r_k$ is the target state vector relative to the observer state $\mathbf{x}^o_{k}$ at time $k$. $\mathbf{F}_k$ is the state transition matrix, and  $\mathbf{w}_k$ is the process noise involved. We assume that the process noise is zero-mean Gaussian $\sim\mathcal{N}\left(\mathbf{w}_k; \mathbf{0}, \mathbf{Q}_k\right)$, with $\mathbf{Q}_k$ as the corresponding covariance matrix. We also assume whiteness of the process noise, which also imparts Markovian property to the system. Hence,
	\begin{align}
		\mathbb{E}\left[\mathbf{w}_l\mathbf{w}_m \right] = \delta_{lm}\mathbf{Q}_l,
	\end{align}
	where $\delta_{lm}$ is the Kronecker delta. The measurement in our case is range-only, which is often corrupted by the additive noise as \cite{bar2001estimation, bhaumik2019nonlinear},
	\begin{align}
		r_k = \norm{\mathbf{H}\mathbf{x}^r_k} + v_k. \label{eq_rangeMeas_additive}
	\end{align}
	Here $\mathbf{H}$ is an appropriate matrix which captures the positional components from $\mathbf{x}_k$, and $v_k$ is the measurement noise, $\sim \mathcal{N}\left(v_k;0,\sigma_r^2\right)$, such that it is white and uncorrelated with $\mathbf{w}_k$. However, for the convenience of our analysis, we use a range measurement model, which is based on adding noise before norm calculation, as asserted in \cite{clark2010gaussian} (discussed in the next subsection).
	
	The objective is to estimate the state $\mathbf{x}_k$ using measurement $r_k$, at each time $k$ such that the solution is optimal in some sense.  
		  
		\subsection{Range Measurement Model}  
		  
		Since range can never be negative, the additive Gaussian noise makes no sense while working in short-range scenarios. For e.g. if the true range is say, $r_k^t$ meters and the standard deviation of the measurement noise is say, $\sigma_r$ meters; then the probability that the resulting range measurement is negative (according to eqn \eqref{eq_rangeMeas_additive}) is, 
		\begin{align}
			\text{Pr}(r_k < 0 ) &= \phi\left( \frac{-r_k^t}{\sigma_r}\right) = \int_{-\infty}^{0}\mathcal{N}\left(r;r_k^t,\sigma_r\right) dr  \notag \\
											&= \frac{1}{2}\left[ 1 - \erf\left(\frac{r_k^t}{\sigma_r\sqrt{2}}\right)\right],
		\end{align}
		where $r^t_k = \norm{\mathbf{H}\mathbf{x}_k} $ is the true range, $\phi(.)$ is the cumulative distribution function (CDF) of a normal distribution, and $\erf(.)$ denotes the error function. Thus, the probability increases with decreasing range and is never theoretically zero. Therefore, we advocate the use of additive Gaussian term inside the norm operator,
			\begin{align}
					r_k = \norm{\mathbf{H}\mathbf{x}_k + \mathbf{v}_k},
			\end{align}
			where $\mathbf{v}_k \sim \mathcal{N}\left(0, \sigma_r^2\mathbf{I}\right)$ with $\mathbf{I}$ being the identity matrix of appropriate dimension. This model, however, gives rise to the following Ricean density for conditional range \cite{rice1944mathematical},
			\begin{align}
				p_r\left(r_k | \mathbf{x}_k\right) = \frac{r_k}{\sigma_r^2}\exp\left( -\frac{r_k^2 + \norm{\mathbf{H}\mathbf{x}_k}^2 }{2\sigma_r^2}\right)I_0\left( \frac{r_k\norm{\mathbf{H}\mathbf{x}_k}}{\sigma_r^2}\right) \label{eq_range_den_rice},
			\end{align}
		where $I_0(.)$ is the modified Bessel function of the first kind and order 0. In contrast, the Gaussian density of the original measurement conditioned on the truth is,
		\begin{align}
			p_g\left(r_k | \mathbf{x}_k\right) = \frac{1}{\sigma_r\sqrt{2\pi}}\exp\left( -\frac{\left(r_k - \norm{\mathbf{H}\mathbf{x}_k}\right)^2 }{2\sigma_r^2}\right).
		\end{align}
		
		It can be observed that as $\frac{r_k}{\sigma_r}$ increases, the difference ${p_r(.)} - {p_g(.)}$ asymptotically converges to ${p_g\left(r_k | \mathbf{x}_k\right)} \left[ \sqrt{\frac{r_k}{r^t_k}} - 1\right]$.

		Therefore, practically, at large range and moderate standard deviations, $r_k \approx r^t_k$, and the densities are indistinguishable. The reason for choosing such a range model is only analytic convenience to arrive at closed-form solutions. To see this, consider linear measurements of position with Gaussian additive noise,
		\begin{align}
			\mathbf{y}_k = r^t_k \begin{bmatrix}
				\cos(\theta_k) \\
				\sin(\theta_k)
			\end{bmatrix} + \mathbf{v}_k \label{eq_linMeas},
		\end{align}
		where $\theta_k$ is the azimuth, measured counter-clockwise from x-axis. It is easier to observe that the density of range measurement, $r_k = \norm{\mathbf{y}_k}$ will coincide with that in Eqn. \eqref{eq_range_den_rice}.
		
	\section{Update Equations for Range-Only Tracking} \label{sec_updateEqn}
		The posterior density of the state $\mathbf{x}_k$ conditioned on range $r_k$ can be represented as the marginalization of azimuth $\theta$ as
		\begin{align}
			p(\mathbf{x}_k|r_k)  =& \int_{0}^{2\pi}  p(\mathbf{x}_k, \theta_k | r_k) d\theta_k \notag \\
								=& \int_{0}^{2\pi} p(\mathbf{x}_k | \theta_k , r_k) p(\theta_k | r_k) d\theta_k \label{exact_rOnly_filtering}
		\end{align}
		The interpretation of the above equations is as follows. Assume that a linear measurement is available as in Eqn. \eqref{eq_linMeas}. Then, the posterior state estimate becomes,
		\begin{align}
			\hat{\mathbf{x}}_{k|k} &= \hat{\mathbf{x}}_{k|k-1} + \mathbf{K}_k\left[\mathbf{y}_k - \hat{\mathbf{y}}_{k} \right] \notag \\
									&= \left( \mathbf{I} - \mathbf{K}_k\mathbf{H}\right)\hat{\mathbf{x}}_{k|k-1} + \mathbf{K}_kr_k\begin{bmatrix}
																																\cos(\theta_k)\\
																																\sin(\theta_k)			
																																\end{bmatrix},
		\end{align}
		where $\hat{\mathbf{y}}_{k} = \mathbf{H}\mathbf{\hat{x}}_{k|k-1}$ is the predicted measurement which can be calculated using the prior density and $\mathbf{K}_k$ is the Kalman gain \cite[Eqn. 5.2.3-11]{bar2001estimation}. Since the azimuth information is not available, we find the posterior estimate by conditioning on range, 
		\begin{align}
			\hat{\mathbf{x}}_{k|k} = \mathbb{E}\left[ {\mathbf{x}}_{k} |r_k \right] = \left( \mathbf{I} - \mathbf{K}_k\mathbf{H}\right)\hat{\mathbf{x}}_{k|k-1} +\mathbf{K}_kr_k \mathbb{E}\left[\mathbf{b}_k|r_k\right] \label{eq_condEst},
		\end{align}
	where, $\mathbf{b}_k = \begin{bmatrix}\cos(\theta_k) & \sin(\theta_k) \end{bmatrix}^T$. Similarly, the corresponding posterior error covariance can be calculated as,
	\begin{align}
		\mathbf{P}_{k|k} = \left( \mathbf{I} - \mathbf{K}_k\mathbf{H}\right) \mathbf{P}_{k|k-1} + r_k^2\mathbf{K}_k\mathbf{P}_{b,k}\mathbf{K}^T_k, 
	\end{align}
	where, the associated covariance due to azimuth uncertainty, $\mathbf{P}_{b,k}$ is quantified as,
	\begin{align}
		\mathbf{P}_{b,k} = \mathbb{E}\left[\left(\mathbf{b}_k - \mathbb{E}[\mathbf{b}_k ]\right)\left(\mathbf{b}_k - \mathbb{E}[\mathbf{b}_k ]\right)^T | r_k \right]. \label{eq_condCov}
	\end{align}
	Though the Kalman filter update equations look promising for generating a near-optimal estimate, there is, however, a missing piece of information which is vital. The conditional expectations in the above equation assert that the azimuth density (conditioned on range), is considered a symmetric uni-modal distribution, which is unfortunately not true in all scenarios. The famous contact-lens problem \cite{tian2014consistency} in range-only tracking poses a conic shaped distribution of posterior state, which has to be dealt with accordingly. The problem has been solved by approximating the unimodal density as a Gaussian mixture, which can be recursively predicted and updated to generate a multimodal posterior density. 
	
	In this work, we focus on deriving the exact expression for the conditional expectation in Eqn. \eqref{eq_condEst} and Eqn. \eqref{eq_condCov} and their generalization. The exact expressions will be helpful in generating accurate multi-modal approximations to the non-trivial conditional density, $p\left(\theta_k|r_k\right)$, as we see in the subsequent sections.
		 
	 \section{Conditional Azimuth Density} \label{sec_condAzimuth_density}
	  The conditional azimuth density can be represented as,
	 \begin{align}
	 	p\left(\theta_k|r_k\right) = \frac{p\left(\theta_k, r_k\right)}{\int\limits_{0}^{2\pi} p\left(\theta_k, r_k\right) d\theta_k  } \label{eq_condAzimuth_density}.
	 \end{align} 
	 Assuming that the true position at time $k$ is Gaussian distributed with predicted positional estimate as the mean, $\mathbf{y}_k \sim \mathcal{N}\left(\mathbf{y}_k; \mathbf{\hat{y}}_k, \mathbf{V}_k \right)$ as in Eqn. \eqref{eq_linMeas}. Where $\hat{y}_k = [\hat{x}_{k|k-1}, \hat{y}_{k|k-1}]^T$ and $(\hat{x}_{k|k-1}, \hat{y}_{k|k-1})$ are the predicted position estimates. Using transformation of variables, the following joint density in Cartesian coordinates,
	 \begin{align}
	 	p(x_k, y_k) = \frac{1}{\sqrt{\abs{2\pi\mathbf{V_k}}}} \exp\left[-\frac{1}{2}\left(\mathbf{y}_k - \mathbf{\hat{y}}_k\right)^T \mathbf{V}^{-1}_k\left(\mathbf{y}_k - \mathbf{\hat{y}}_k\right)\right]
	 \end{align}
	 can be transformed to polar coordinates,
	 \begin{align}
	 	p(\theta_k, r_k) =& \frac{r_k}{\sqrt{|2\pi \mathbf{V}_k|}}\exp\left[-\frac{1}{2}\left(r_k^2 \mathbf{b}_k^T\mathbf{V}_k^{-1}\mathbf{b}_k + \hat{\mathbf{y}}_k^T\mathbf{V}_k^{-1}\hat{\mathbf{y}}_k \right)\right] \notag \\
	 					&\times \exp\left[-\frac{1}{2} \left( -2r_k\mathbf{b}^T_k\mathbf{V}_k^{-1}\hat{\mathbf{y}}_k \right)  \right],
	 \end{align}
 	where $\mathbf{V}_k = \mathbf{H}\mathbf{P}_{k|k-1}\mathbf{H}^T + \mathbf{R}_k$ is the predicted measurement covariance. The joint density can be rewritten as,
 \begin{align}
 		p&(\theta_k, r_k) = \frac{r_k}{\sqrt{|2\pi \mathbf{V}_k|}}\exp\left[-\frac{1}{2}\left(r_k^2\left(\frac{a}{2} + \frac{c}{2}\right) + 								\hat{\mathbf{y}}_k^T\mathbf{V}_k^{-1}\hat{\mathbf{y}}_k \right)\right] \notag \\
 		&\times \exp\biggl[r_k \norm{p,q} \cos(\theta_k-\phi_1) + A_3r_k^2 \cos(2\theta_k + \phi_2)\biggr] \notag \\
 		&= \kappa_k \exp\biggl[r_k \norm{p,q} \cos(\theta_k-\phi_1) + A_3r_k^2 \cos(2\theta_k + \phi_2)\biggr],
 	\end{align}
 	where $\kappa_k$ contains all the term independent of $\theta_k$,
 	\begin{align}
 		\kappa_k = \frac{r_k}{\sqrt{|2\pi \mathbf{V}_k|}}\exp\left[-\frac{1}{2}\left(r_k^2\left(\frac{a}{2} + \frac{c}{2}\right) + 								\hat{\mathbf{y}}_k^T\mathbf{V}_k^{-1}\hat{\mathbf{y}}_k \right)\right], \label{eqn_kappa}
 	\end{align}
 	and the rest of the notations employed are defined as follows,
	\begin{align}
		 \begin{bmatrix}
							a & b\\
							b & c
							\end{bmatrix} = \mathbf{V}_k^{-1}, \quad  \begin{bmatrix}
																			p\\q
																		\end{bmatrix} &= \mathbf{V}_k^{-1}\hat{\mathbf{y}}_k, \quad \phi_1 = \tan^{-1}\left(\frac{q}{p}\right),\notag \\
		A_3 = \norm{\left(\frac{c}{4} - \frac{a}{4}\right), \frac{b}{2}}, \quad \phi_2 &= \tan^{-1}\left(\frac{b/2}{\left(\frac{c}{4} - \frac{a}{4}\right)}\right).\label{eq_param_1}
	\end{align}
	Note that the notation $\norm{\alpha, \beta}$ means the norm of a vector containing $\alpha$ and $\beta$.
	
	\subsection{True Range Density}
	The true range density can be computed by solving the integral
	\begin{align}
		p(r_k) &= \int\limits_{0}^{2\pi} p\left(\theta_k, r_k\right) d\theta_k = \kappa_k \int_0^{2\pi} \exp\bigl[r_k \norm{p,q} \times \notag \\ 
										&\quad \cos(\theta_k-\phi_1) + A_3r_k^2 \cos(2\theta_k + \phi_2)\bigr] d\theta_k, \label{eq_rangeDensity_integral}
	\end{align}
	where $\kappa_k$ is defined in Eqn. \eqref{eqn_kappa}. The integral has been solved by Weil for the diagonal case in \cite{weil1954distribution}. For the general case of Eqn. \eqref{eq_rangeDensity_integral}, the integral can be solved as a series expansion (see proof in appendix) wherein the density can be calculated by a relatively fewer number of terms.  The resulting range density which is also the denominator in Eqn. \eqref{eq_condAzimuth_density} is,
	 \begin{align}
		p(r_k) &= \kappa(r_k) 2\pi \biggl[ I_0\left(A_3r_k^2\right)I_0\left(r_k\sqrt{A_1^2 + A_2^2}\right) \notag \\
				 &+ 2\sum_{k=1}^{\infty} I_k\left(A_3r_k^2\right)I_{2k}\left(r_k\sqrt{A_1^2 + A_2^2}\right)\cos(2k\psi) \biggr]\label{eq_trueRange}
	\end{align}
	where $I_v(.)$ is the modified Bessel function of the first kind of order $v$, and,
	\begin{align}
		A_1 = \norm{p,q} \cos(\frac{\phi_2}{2} + \phi_1)&, \quad A_2 = \norm{p,q} \sin(\frac{\phi_2}{2}+ \phi_1),\notag \\
			 \psi =& \tan^{-1}\left(\frac{A_2}{A_1}\right). \label{eq_const_A1_A2}
	\end{align}
	The resulting conditional azimuth density is then,
	\begin{align}
			p\left(\theta_k|r_k\right) = \frac{\kappa_k}{p\left(r_k\right)} \times \exp\biggl[r_k \norm{p,q} \cos(\theta_k-\phi_1)\quad+ \notag \\ A_3r_k^2 \cos(2\theta_k + \phi_2)\biggr] \label{eq_condAzimDensity},
	\end{align}
	\noindent where $p(r_k)$ is given in Eqn. \eqref{eq_trueRange}, note that $\kappa_k$ will be canceled out. The remaining denominator in Eqn. \eqref{eq_condAzimDensity} is
	\begin{align}
		&\Delta_k = 2\pi \biggl[ I_0\left(A_3r_k^2\right)I_0\left(r_k\sqrt{A_1^2 + A_2^2}\right) + \notag \\
		&\quad + 2\sum_{k=1}^{\infty} I_k\left(A_3r_k^2\right)I_{2k}\left(r_k\sqrt{A_1^2 + A_2^2}\right)\cos(2k\psi) \biggr],
	\end{align}
	which can be approximated by a fewer number of terms in summation.
	
	\section{Expectations of Conditional Azimuth Density} \label{sec_expOfCond_density}
	To calculate the optimal solution in eqns. \eqref{eq_condEst} and \eqref{eq_condCov}, it is necessary to calculate the expectations of the form $\mathbb{E}\left[\cos(\theta_k)|r_k\right]$, $\mathbb{E}\left[\sin(\theta_k)|r_k\right]$, and others. We show that it is possible to compute the generalized expectations of the form $\mathbb{E}\left[\cos(m\theta_k)|r_k\right]$ and $\mathbb{E}\left[\sin(m\theta_k)|r_k\right]$ for any positive integer $m$. 		
		\begin{align}
		\mathrm{I}_\text{gen,cos}(m)&=\mathbb{E}\left[\cos(m\theta_k)|r_k\right] =  \int_0^{2\pi} \cos(m\theta_k) p\left(\theta_k|r_k\right)d\theta_k,  \notag										
		\end{align}
		or,
		\begin{align}
			\mathrm{I}_\text{gen,cos}(m) &\propto \int_0^{2\pi} \cos(m\theta_k) \exp\bigl[r_k \norm{p,q} \cos(\theta_k-\phi_1)  \notag\\ 
			& \qquad \qquad + A_3r_k^2 \cos(2\theta_k + \phi_2)\bigr] d\theta_k \label{eq_genInteral_cos},
		\end{align}
		where the proportionality constant is $1/\Delta_k$. In order to proceed with the integral, let's solve a simplified version for the case where the covariance matrix $\mathbf{V}_k$ is diagonal,
			\begin{align}
			 \mathrm{I}_\text{diag,cos}(m) =\int_0^{2\pi}\cos(m\theta_k)\exp&\bigl[r_k \norm{p,q} \cos(\theta_k-\phi_1) +\notag \\
																		 &A_3r_k^2 \cos(2\theta_k)\bigr] d\theta_k, \label{eq_diagIntegral_cos}
		\end{align}
		and, 
		\begin{align}
				\mathrm{I}_\text{diag,sin}(m) = \int_0^{2\pi}\sin(m\theta_k)\exp&\bigl[r_k \norm{p,q} \cos(\theta_k-\phi_1) +\notag \\
				&A_3r_k^2 \cos(2\theta_k)\bigr] d\theta_k. \label{eq_diagIntegral_sin}
		\end{align}		
		 Note that the difference between the generalized integrals and the integrals for the diagonal case is in the inclusion of $\phi_2$. 
		 It is proven in the appendix that Eqns. \eqref{eq_diagIntegral_cos} and \eqref{eq_diagIntegral_sin} can further be reduced to,
		\begin{align}
			\mathrm{I}_\text{diag,cos}(m) = 2\pi\sum_{j = -\infty}^\infty &\frac{1}{2} I_j(A_3r_k^2) \bigl[I_{2j + m}(Dr_k)\cos((2j + m)\psi) \notag \\
																		&+ I_{2j - m}(Dr_k)\cos((2j - m)\psi) \bigr], \label{eq_diagCos}
		\end{align}
		and,
		\begin{align}
			\mathrm{I}_\text{diag,sin}(m) = 2\pi\sum_{j = -\infty}^\infty &\frac{1}{2} I_j(A_3r_k^2) \bigr[I_{2j + m}(Dr_k)\sin((2j + m)\psi) \notag \\
																		&- I_{2j - m}(Dr_k)\sin((2j - m)\psi) \bigl], \label{eq_diagSin}
		\end{align}
		where $D = \sqrt{(A_1^2 + A_2^2)}$ and $\tan(\psi) = \frac{A_2}{A_1}$.
		Moving forward, note that the term inside the summation is an even function of summation index $j$. Therefore, the infinite series in Eqn. \eqref{eq_diagCos} and \eqref{eq_diagSin} can be reduced to the following expressions,
		\begin{align}
			&\mathrm{I}_\text{diag,cos}(m) = 2\pi \bigl[I_0(A_3r_k^2) I_{m}(Dr_k)\cos(m) + \sum_{j = 1}^\infty I_j(A_3r_k^2) \times \notag \\
			& \bigl(I_{2j + m}(Dr_k)\cos((2j + m)\psi)+ I_{2j - m}(Dr_k)\cos((2j - m)\psi) \bigr) \bigr], \label{eq_diagCos_pos}\\
			&\mathrm{I}_\text{diag,sin}(m) = 2\pi \bigl[I_0(A_3r_k^2) I_{m}(Dr_k)\sin(m) + \sum_{j = 1}^\infty I_j(A_3r_k^2) \times \notag \\
			& \bigl(I_{2j + m}(Dr_k)\sin((2j + m)\psi) - I_{2j - m}(Dr_k)\sin((2j - m)\psi) \bigr) \bigr], \label{eq_diagSin_pos} \notag \\
		\end{align}
		which reduces the computation time by half.
		
		The intergal for the general case in Eqn. \eqref{eq_genInteral_cos} can be converted to that in eqns. \eqref{eq_diagIntegral_cos} and \eqref{eq_diagIntegral_sin} using the substitution $2\theta_k + \phi_2 = 2u$. Which results in the following expressions for the integrals of our interest,
		\begin{align}
			\mathrm{I}&_\text{gen,cos}(m) = \mathbb{E}[\cos(m\theta_k)]_{p(\theta_k|r_k)} \notag \\
			&= \frac{2\pi}{\Delta_k} \biggl[\cos(m\frac{\phi_2}{2})\mathrm{I}_\text{diag,cos} +  \sin(m\frac{\phi_2}{2})\mathrm{I}_\text{diag,sin}\biggr], \label{eq_IntGen_nonsimp_cos}\\
			\mathrm{I}&_\text{gen,sin}(m) = \mathbb{E}[\sin(m\theta_k)]_{p(\theta_k|r_k)} \notag \\
			 &= \frac{2\pi}{\Delta_k} \left[\cos(m\frac{\phi_2}{2})\mathrm{I}_\text{diag,sin} -  \sin(m\frac{\phi_2}{2})\mathrm{I}_\text{diag,cos}\right],\label{eq_IntGen_nonsimp_sin}
		\end{align}
		where $\mathrm{I}_\text{diag,cos}$ and $\mathrm{I}_\text{diag,sin}$ are given in eqns. \eqref{eq_diagCos_pos} and \eqref{eq_diagSin_pos} respectively. 
		
		Since $r_k$ is generally of the order of thousands of meters, attempting to calculate the Bessel functions in above expressions directly might result in floating point errors. One way is to use the following approximation for the ratio of Bessel functions with the same argument \cite{olver2010nist,weisstein2012bessel, NIST:DLMF},
		\begin{align}
			\frac{I_N(x)}{I_0(x)} \approx \exp(-\frac{N^2}{2x}) \qquad x >> N^2 >> 1. \label{eq_besselRatio_approx}
		\end{align}
		 The approximation tends to get accurate with increasing argument $x$. Under this assumption, the error was found to be of the order $\mathcal{O}\left(\frac{N^4}{x^2}\right)$ \cite{1495410}.
		
		Dividing the numerator and denominator in eqns. \eqref{eq_IntGen_nonsimp_cos} and \eqref{eq_IntGen_nonsimp_sin} by $I_0\left(A_3r_k^2\right)I_0\left(r_kD\right)$, and using Eqn. \eqref{eq_besselRatio_approx}, the result is,
		\begin{align}
			\mathrm{I}_\text{gen,cos}(m) = \frac{\cos(m\frac{\phi_2}{2})\mathrm{I}^\text{n}_\text{diag,cos}(m) +  \sin(m\frac{\phi_2}{2})\mathrm{I}^\text{n}_\text{diag,sin}(m)}{1 + 2\sum\limits_{k=1}^{\infty} \exp(-\frac{k^2}{2A_3r^2})\exp(-\frac{(2k)^2}{Dr_k})\cos(2k\psi)} \\
			\mathrm{I}_\text{gen,sin}(m) = \frac{\cos(m\frac{\phi_2}{2})\mathrm{I}^\text{n}_\text{diag,sin}(m) -  \sin(m\frac{\phi_2}{2})\mathrm{I}^\text{n}_\text{diag,cos}(m)}{1 + 2\sum\limits_{k=1}^{\infty} \exp(-\frac{k^2}{2A_3r^2})\exp(-\frac{(2k)^2}{Dr_k})\cos(2k\psi)}
		\end{align}
		where $\mathrm{I}^\text{n}_\text{diag,cos} (\mathrm{I}^\text{n}_\text{diag,sin})$ is same as $\mathrm{I}_\text{diag,cos} (\mathrm{I}_\text{diag,sin})$ normalized with $I_0\left(A_3r_k^2\right)I_0\left(r_kD\right)$, given by the following expression,
		\begin{align}
			&\mathrm{I}^\text{n}_\text{diag,cos}(m) = \frac{\mathrm{I}_\text{diag,cos}(m)}{I_0\left(A_3r_k^2\right)I_0\left(r_kD\right)}  \notag \\
			&= \sum_{j = 0}^\infty
			\exp\left(-\frac{j^2}{2A_3r_k^2}\right) \bigl[\exp\left(-\frac{(2j+m)^2}{2Dr_k}\right)\cos(2j + m) \notag \\
			&\qquad+\exp\left(-\frac{(2j-m)^2}{2Dr_k}\right)\cos(2j - m) \bigr],
		\end{align}
		\begin{align}
			&\mathrm{I}^\text{n}_\text{diag,sin}(m) = \frac{\mathrm{I}_\text{diag,sin}(m)}{I_0\left(A_3r_k^2\right)I_0\left(r_kD\right)}  \notag \\
			&= \sum_{j = 0}^\infty\exp\left(-\frac{j^2}{2A_3r_k^2}\right)\bigl[\exp\left(-\frac{(2j+m)^2}{2Dr_k}\right)\sin(2j + m) \notag \\
			&\qquad -\exp\left(-\frac{(2j-m)^2}{2Dr_k}\right)\sin(2j - m) \bigr].		
		\end{align}
		Note that the condition $x >> N^2 >> 1$ is crucial for the approximation to be meaningful. Other ways for computing Bessel function ratios are mentioned in \cite{amos1974computation} or, directly by using a high precision computing mechanism like Matlab's\textsuperscript{\textcopyright} \texttt{vpa}.
		
		\subsection{Trigonometric Moments}
	 	Once the generalized expectations are formulated, the covariance and higher-order moments can be calculated using known trigonometric identities. For instances, the mean is given by,
		\begin{align}
		\mathbb{E}[\mathbf{b}_k|r_k] = \mathbb{E}\left[\begin{array}{c|c}
					\cos(\theta_k) & \multirow{2}{*}{$r_k$}\\
					\sin(\theta_k) &
			\end{array}\right] = \begin{bmatrix}
									\mathrm{I}_\text{gen,cos}(1) \\
									\mathrm{I}_\text{gen,sin}(1)
								\end{bmatrix}.
		\end{align}
		The covariance $\mathbf{P}_{b,k}$ in Eqn. \eqref{eq_condCov} is given by the formula,
		\begin{align}
			\mathbf{P}_{b,k} &=\mathbb{E}\left[\begin{array}{cc|c}
									\cos^2(\theta_k) & \cos(\theta)\sin(\theta_k) & \multirow{2}{*}{$r_k$}\\
									\sin(\theta_k)\cos(\theta_k) & \sin^2(\theta_k)
									\end{array}		
								\right] \notag \\
							&\qquad - \mathbb{E}[\mathbf{b}_k]\mathbb{E}[\mathbf{b}_k]^T. \label{eq_cov_bearing}
		\end{align}
		where the expectations can be calculated element-wise.
		\begin{align}
			\mathbb{E}\left[\cos^2(\theta_k)|r_k\right] &= \frac{1}{2} + \frac{\mathrm{I}_\text{gen,cos}(2)}{2},\\
			 \mathbb{E}\left[\sin^2(\theta_k)|r_k\right] &= \frac{1}{2} - \frac{\mathrm{I}_\text{gen,cos}(2)}{2},  \\
			\mathbb{E}\left[\cos(\theta_k)\sin(\theta_k)|r_k\right] &= \frac{\mathrm{I}_\text{gen,sin}(2)}{2}.		
		\end{align}
		It is even possible to calculate higher order powered moments using power reduction trigonometric identities. For instance,
		\begin{align}
			\mathbb{E}\left[\cos^3(\theta_k)|r_k\right] &= \frac{3\mathrm{I}_\text{gen,cos}(1) + \mathrm{I}_\text{gen,cos}(3)}{4}, \\
			\mathbb{E}\left[\cos^4(\theta_k)|r_k\right] &= \frac{3 + 4\mathrm{I}_\text{gen,cos}(2) + \mathrm{I}_\text{gen,cos}(4)}{8}, \\
			\mathbb{E}\left[\sin^3(\theta_k)|r_k\right] &= \frac{3\mathrm{I}_\text{gen,sin}(1) - \mathrm{I}_\text{gen,sin}(3)}{4}, \\
			\mathbb{E}\left[\sin^4(\theta_k)|r_k\right] &= \frac{3 - 4\mathrm{I}_\text{gen,cos}(2) + \mathrm{I}_\text{gen,cos}(4)}{8}.
		\end{align}
		Thus, once the recursive functions, $\mathrm{I}_\text{gen,cos}(m)$ and $\mathrm{I}_\text{gen,sin}(m)$ are constructed, it is trivial to calculate any higher order moment.

		\section{Deterministic Sampling of Conditional Azimuth Density and Application to Tracking}\label{sec_detSample}
		
		In the previous section, we proved that the azimuth density can be computed in closed form and their trigonometric moments (also read as circular moments) exist as infinite series which can be approximated well using a few number of terms (see simulation). In range-only filtering, the Gaussian approximation is seldom useful due to the contact-lens problem and thus, it is essential to capture the multi-modality of conditional azimuth density.
		
		A viable solution to this difficulty is to approximate the conditional azimuth density using its deterministic samples and then proceed with Eqn. \eqref{exact_rOnly_filtering} to get the posterior estimate of the target state. Deterministic samples from circular moments have been previously proposed in \cite{kurz2016methods, hanebeck2014moment, clark2010gaussian} but we take a slightly straightforward optimization based approach here which is explained as follows.
		
		The wrapped Dirac distribution is a perfect candidate for the deterministic sampling procedure. The distribution is given by,
		\begin{align}
			\mathit{p}_d(\theta) = \sum_{l=1}^L \gamma_l \delta(\theta - \hat{\theta}_l)
		\end{align} 
		where $\hat{\theta}_l \in [0, 2\pi]$ are the sigma-points position and $\gamma_l \in [0,1]$ are the sigma-points weights corresponding to $\theta_l$ such that $\sum_l \gamma_l = 1$. Thus, there are $2L$ variable to be calculated for Dirac distribution consisting of $L$ components. The $m^\text{th}$ trigonometric moments of a wrapped Dirac distribution are,
		\begin{align}
			\mathbb{E}[\cos(m\theta)]_{\mathit{p}_d(\theta)} &= \sum_{l=1}^L \gamma_l\cos(m\hat{\theta}_l), \\
			\mathbb{E}[\sin(m\theta)]_{\mathit{p}_d(\theta)} &= \sum_{l=1}^L \gamma_l\sin(m\hat{\theta}_l)
		\end{align}
		 
		For approximation of $\mathit{p}(\theta_k|r_k)$, we substitute the L.H.S in the above equations with our series-approximated $m^\text{th}$ order moments and solve for $\gamma = \left[\gamma_1, \gamma_2, \cdots, \gamma_L\right]$ and $\theta = \left[\theta_1, \theta_2, \cdots, \theta_L\right]$. Thus, a total of $2M$ equations need to be solved for an $L-$component distribution, where $M$ is the highest order considered for computation. Mathematically, we solved the following norm-minimization problem to arrive at optimal values of $\gamma$ and $\theta$.
		\begin{align}
			\gamma^*, \theta^* = \argmin_{\gamma, \theta} \norm{\begin{matrix} \\ I_\text{gen,cos}(1) - \sum\limits_{l=1}^L \gamma_l\cos(\hat{\theta}_l) \\
			I_\text{gen,sin}(1) -  \sum\limits_{l=1}^L \gamma_l\sin(\hat{\theta}_l)\\\vdots\\ I_\text{gen,cos}(M) - \sum\limits_{l=1}^L \gamma_l\cos(\hat{M\theta}_l) \\
			I_\text{gen,sin}(M) -  \sum\limits_{l=1}^L \gamma_l\sin(M\hat{\theta}_l)\end{matrix} },  \label{eq_optProb}
		\end{align}
		\begin{align}
			&\text{such that,} \notag \\
			& \quad 1 \geq \gamma_l \geq 0; \quad 2\pi \geq \theta_l \geq 0 \quad \forall l = 1,2, \cdots, L \notag
		\end{align}
		The optimal values of $\gamma$ obtained from above are then normalized such that $\sum_l \gamma^*_l = 1$. 
		
\begin{figure}[t]
			\centering
			\begin{subfigure}{0.32\linewidth}
				\includegraphics[width=\columnwidth]{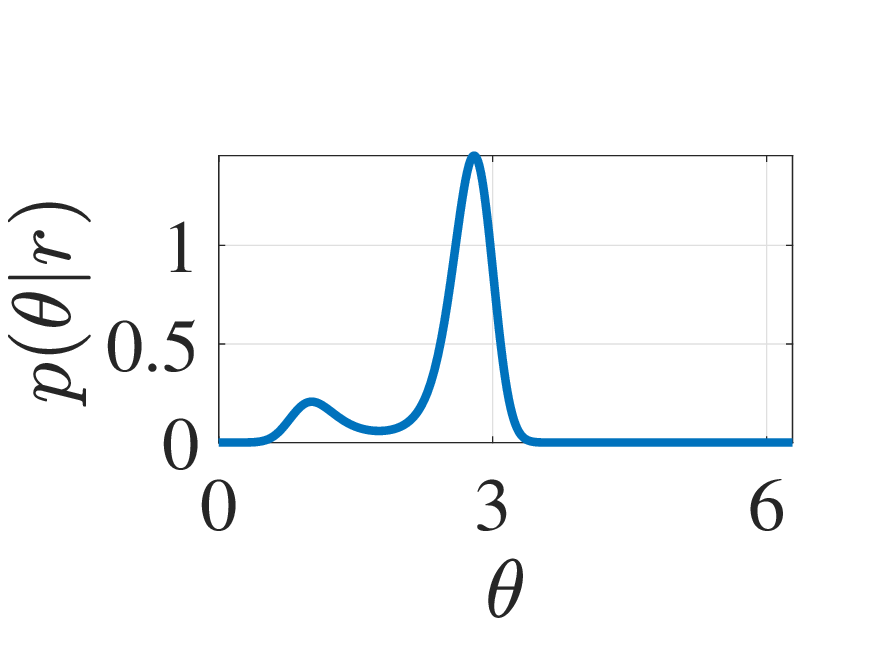}
			\end{subfigure}
			\begin{subfigure}{0.32\linewidth}
				\includegraphics[width=\columnwidth]{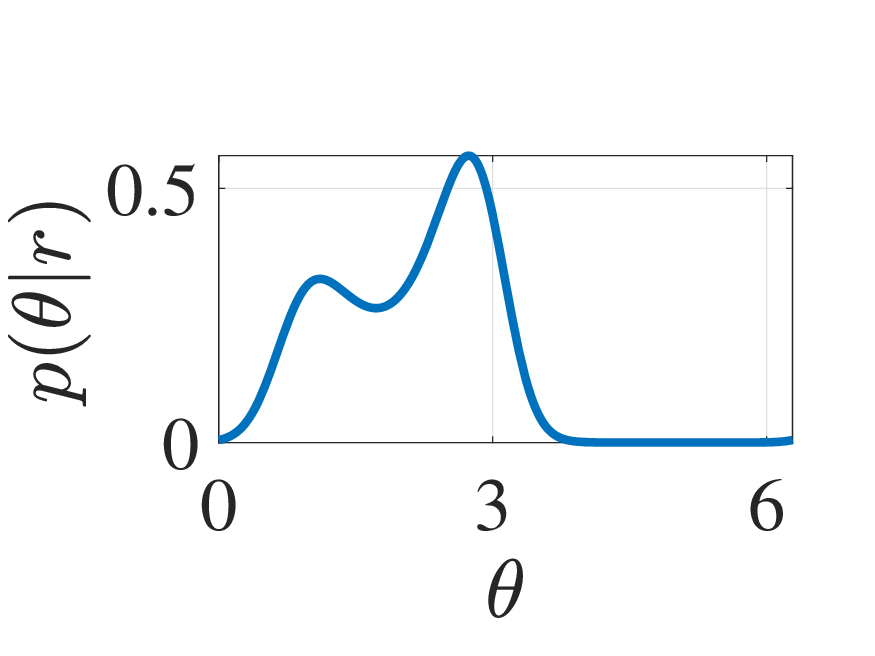}
			\end{subfigure}
			\begin{subfigure}{0.32\linewidth}
				\includegraphics[width=\columnwidth]{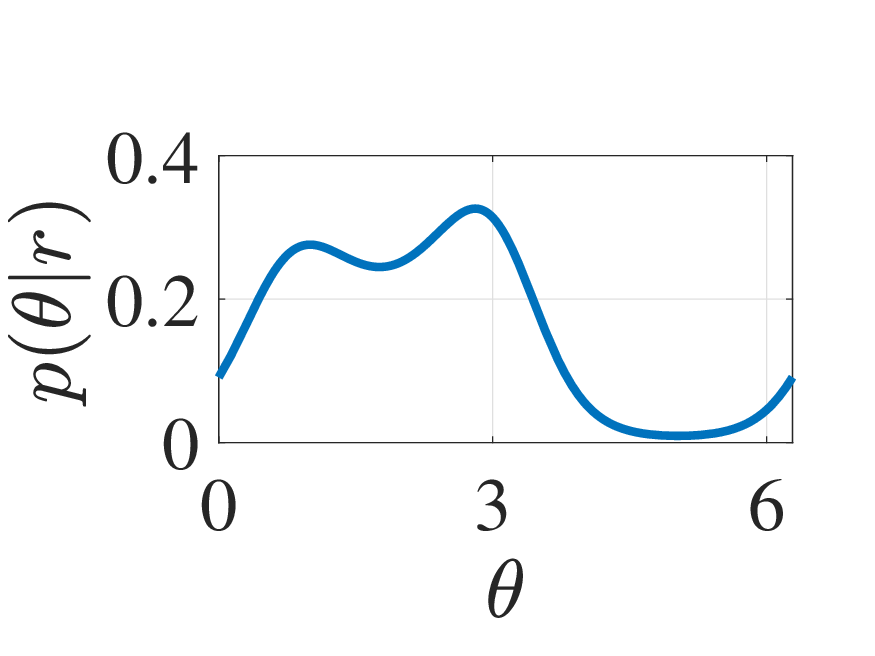}
			\end{subfigure} \par\medskip
			
			\begin{subfigure}{0.32\linewidth}
				\includegraphics[width=\columnwidth]{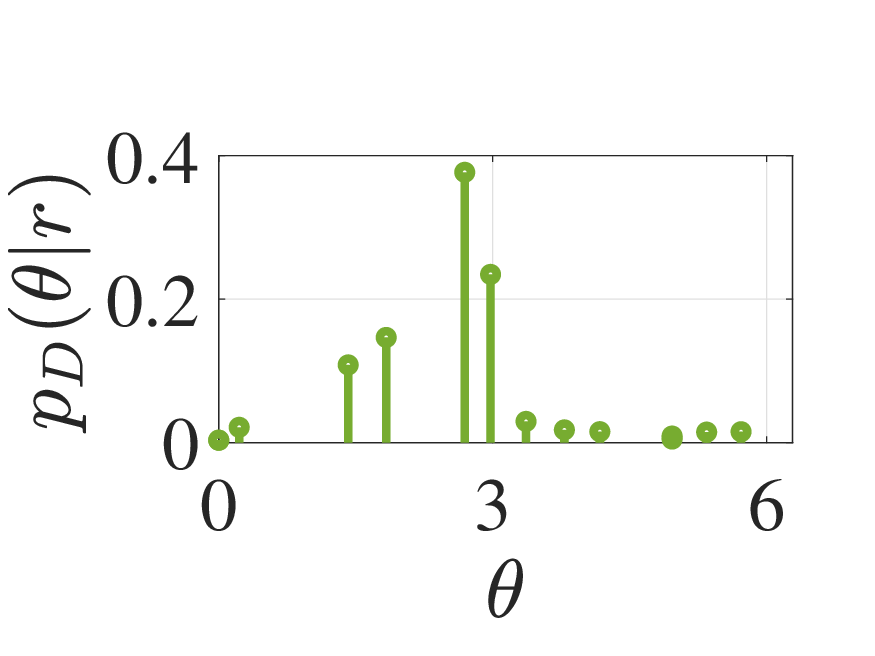}
			\end{subfigure}
			\begin{subfigure}{0.32\linewidth}
				\includegraphics[width=\columnwidth]{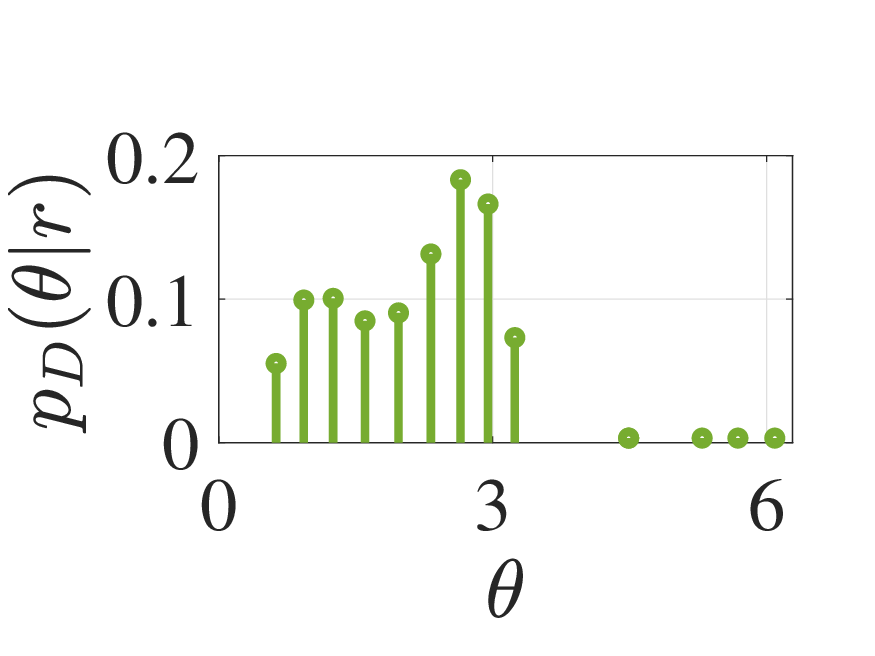}
			\end{subfigure}
			\begin{subfigure}{0.32\linewidth}
				\includegraphics[width=\columnwidth]{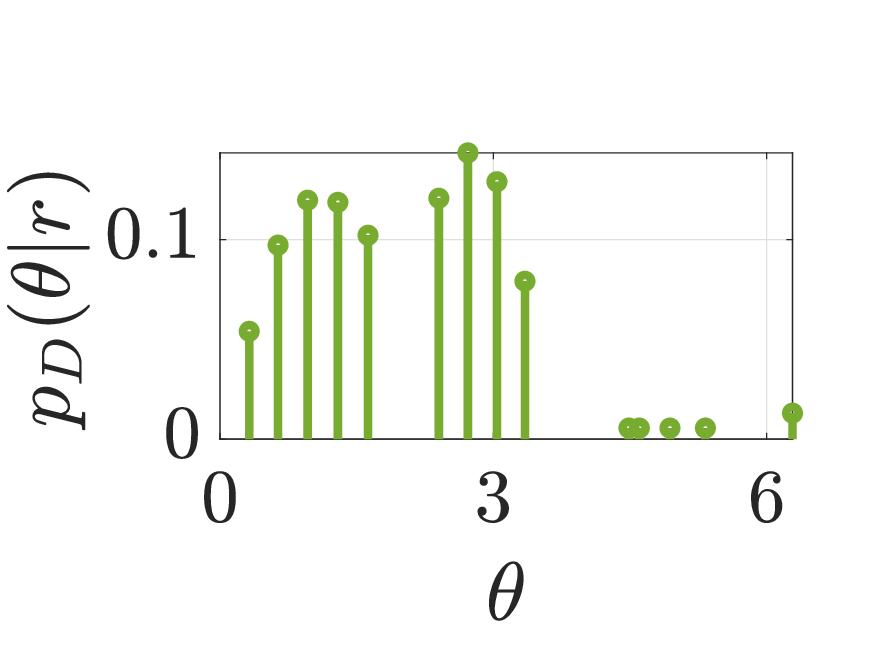}
			\end{subfigure}\par \medskip
			
			\begin{subfigure}{0.32\linewidth}
				\includegraphics[width=\columnwidth]{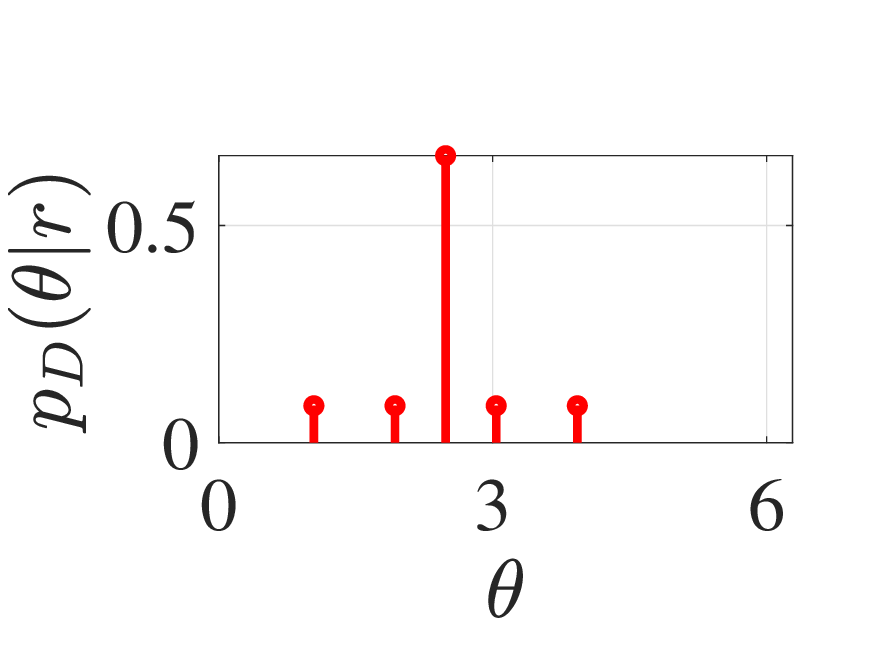}
				\caption*{$q = 0.25$.}
			\end{subfigure}
			\begin{subfigure}{0.32\linewidth}
				\includegraphics[width=\columnwidth]{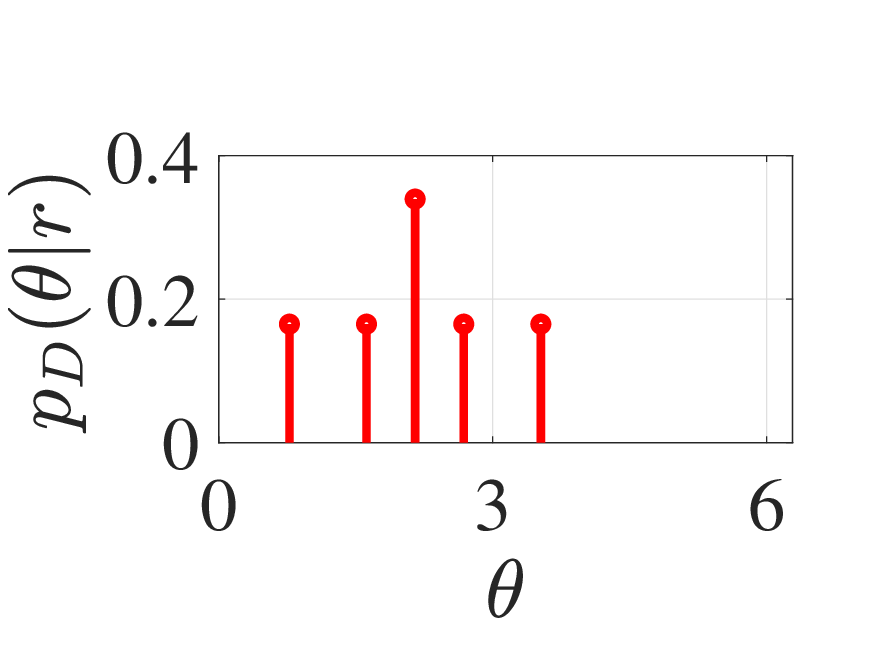}
				\caption*{$q = 0.75.$}
			\end{subfigure}
			\begin{subfigure}{0.32\linewidth}
				\includegraphics[width=\columnwidth]{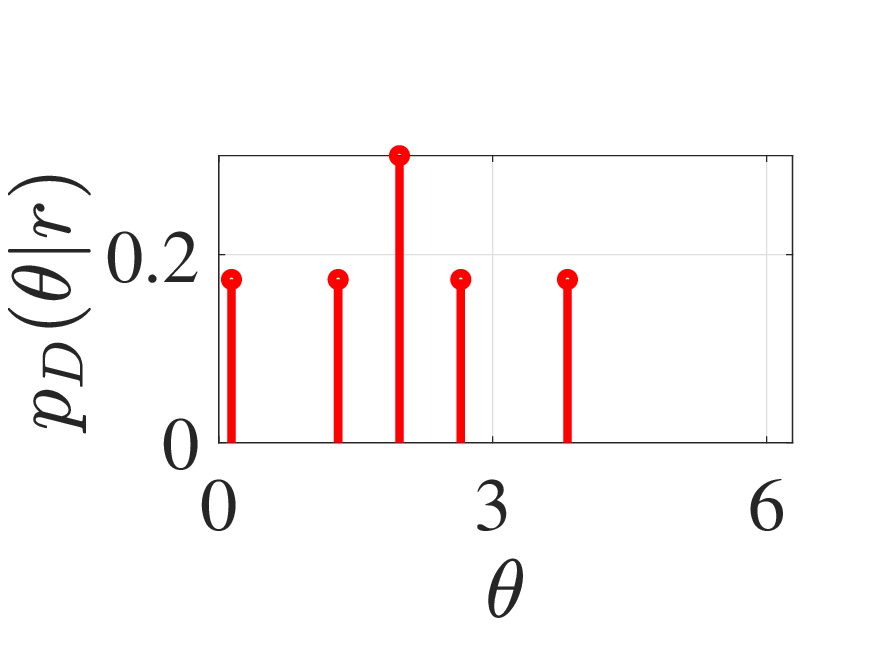}
				\caption*{$q = 3.$}
			\end{subfigure}
			\caption{(Top) Actual Density, (Middle) Deterministic Sampling using proposed method, (Bottom) Using the technique specified in \cite{kurz2016methods}; for various values of $q$. }
			\label{fig_determSampling}
		\end{figure}

		For demonstration, we plotted the conditional azimuth density and the corresponding values of sigma-points $\gamma, \hat{\theta}$ for $\mathbf{V}_k = q\times10^3\begin{bmatrix}
			7 & 2\\
			2 & 1
		\end{bmatrix}$ and $\hat{\mathbf{y}}_k= \begin{bmatrix}
		-50\\20			
		\end{bmatrix}$ in Fig. \ref{fig_determSampling} for three different values of $q$.
		
		Where $q$ is any positive real number. For comparison, the symmetric sampling technique, which uses only the first and second order moments \cite{kurz2016methods}, is also plotted in Fig. \ref{fig_determSampling}. The optimization was based on 14 components in the Dirac distribution using the first 7 moments (7 moments for $\sin$ \& $\cos$ respectively, thus equating 14 components). It can be seen that the optimization based approach faithfully captures the multi-modality and skewness of the original distribution. More components can be added for accuracy at the cost of computational time.
		
		Matlab's\textsuperscript{\textcopyright} \texttt{fmincon} was used to minimize Eqn. \eqref{eq_optProb} using the sequential quadratic programming (SQP) algorithm \cite{nocedal1999numerical}. The algorithm found an optima much faster when the system was over-determined, which means that the number of moments (M) employed were greater than the number of component parameters (L).

		\subsection{Proposed Algorithm for 2D Range-Only Tracking}
		Assuming that a Gaussian prior $\hat{\mathbf{x}}_{k|k-1}, \mathbf{P}_{k|k-1}$ is available, the proposed algorithm is as follows :

		\begin{itemize}
		\item Choose an integer $N$ for the number of terms used in the summation of Eqn. \eqref{eq_diagCos_pos} and \eqref{eq_diagSin_pos}. 
		\item Choose an integer $M$ for the order of circular moments to be computed.
		\item Calculate the parameters $\mathbf{V}_k, p, q, \phi_1, A_1, A_2, A_3$ in eqns. \eqref{eq_param_1} and \eqref{eq_const_A1_A2}.
		\item Compute circular moments, $I_\text{gen,cos}(m)$ and $I_\text{gen,sin}(m)$ for $m = 1,\cdots,M$ from eqns. \eqref{eq_IntGen_nonsimp_cos} and \eqref{eq_IntGen_nonsimp_sin}.
		\item From circular moments, calculate $L$ deterministic points $(\hat{\theta}_{l,k})$ and corresponding weights $\gamma_{l}$ of the conditional azimuth density by solving the optimization problem in Eqn. \eqref{eq_optProb}.
		\item Substitute the Dirac approximation in Eqn. \eqref{exact_rOnly_filtering},
		\begin{align}
			p(\mathbf{x}_k|r_k)  =& \int_{0}^{2\pi} p(\mathbf{x}_k | \theta_k , r_k) p(\theta_k | r_k) d\theta_k \notag \\
			=& 	\int_0^{2\pi} p(\mathbf{x}_k | \theta_k , r_k) \sum_{l=1}^L \gamma_l \delta(\theta_k - \hat{\theta}_{l,k})d\theta_k \notag \\
			=& \sum_{l=1}^L \gamma_l \mathcal{N}\left(\mathbf{x}_k; \hat{\mathbf{x}}^l_{k|k}, \mathbf{P}^l_{k|k} \right) \label{eq_gm_posterior}
		\end{align}
		where the local components $\hat{\mathbf{x}}^l_k, \mathbf{P}^l_k$ are given by,
		\begin{align}
			\hat{\mathbf{x}}^l_{k|k} &= \left(\mathbf{I} - \mathbf{K}_k\mathbf{H}\right)\hat{\mathbf{x}}_{k|k-1}  + r_k\mathbf{K}_k\begin{bmatrix}\cos(\hat{\theta}_{l,k}) \\ \sin(\hat{\theta}_{l,k}) \end{bmatrix} \label{eq_PosteriorMode_estimate} \\
			\mathbf{P}^l_{k|k} &= \left(\mathbf{I} - \mathbf{K}_k\mathbf{H}\right)\mathbf{P}_{k|k-1} + r_k^2\mathbf{K}_k \mathbf{P}_{b,k}\mathbf{K}^T_k
		\end{align}
		where the covariance $\mathbf{P}_{b,k}$ is given in Eqn. \eqref{eq_cov_bearing}.
		\item As expected, the resulting posterior in Eqn. \eqref{eq_gm_posterior} takes the form of a Gaussian mixture. Find the resulting mean and covariance, and propagate for the next recursion. 
		\end{itemize}
		It is obvious that the algorithm can be made more robust by using the Gaussian mixture itself as a prior instead of its Gaussian approximation. That will, however, have an impact on the computational requirements of the algorithm since mixture reduction techniques would have to be applied too. Thus, the authors intend to employ it when efficient techniques for calculating ratio of Bessel functions and optimization in Eqn. \eqref{eq_optProb} have been developed. As a proof-of-concept, the simulation uses a Gaussian prior for all filtering recursions.
		
		\begin{figure*}[t]
			\centering
			\begin{subfigure}{0.24\linewidth}
				\centering
				\includegraphics[width=\columnwidth]{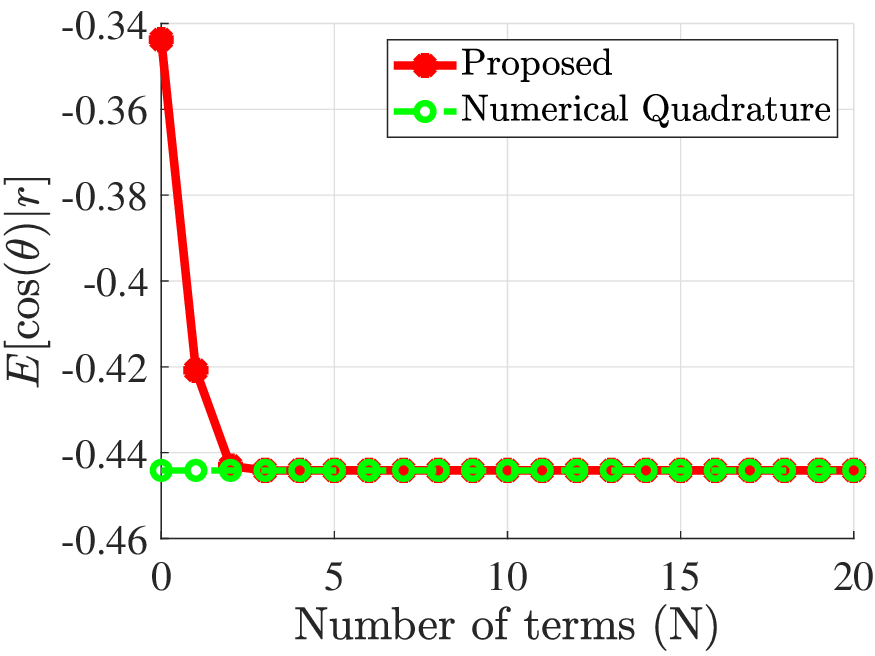}
				\caption{$m = 1$.}
			\end{subfigure}
			\begin{subfigure}{0.24\linewidth}
				\centering
				\includegraphics[width=\columnwidth]{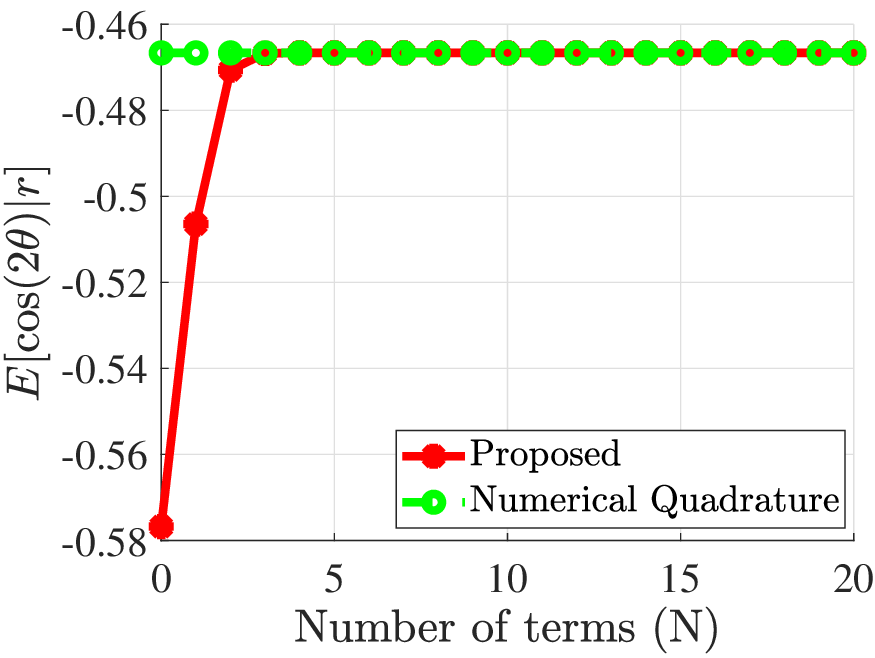}
				\caption{$m = 2$.}
			\end{subfigure}
			\begin{subfigure}{0.24\linewidth}
				\centering
				\includegraphics[width=\columnwidth]{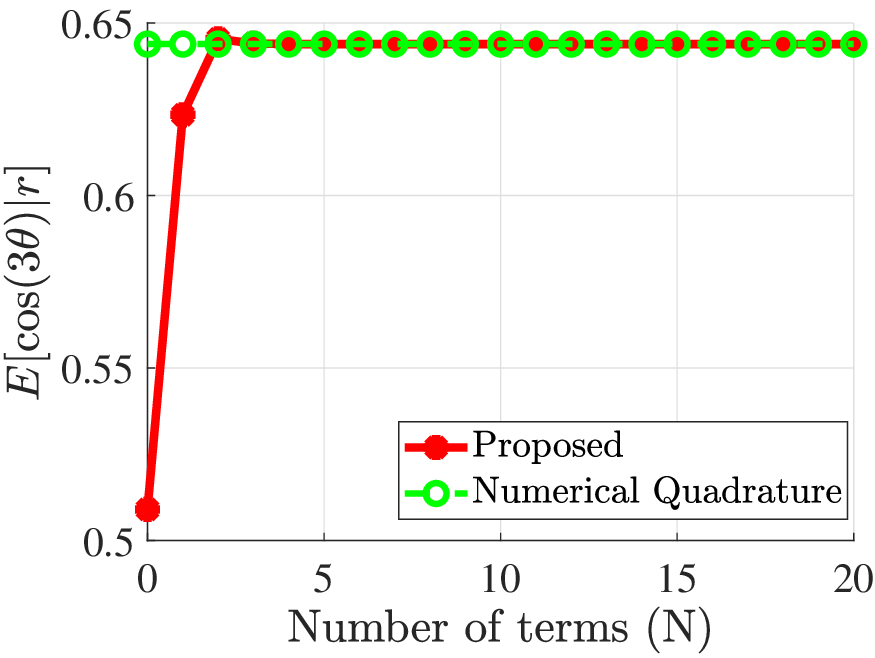}
				\caption{$m = 3$.}
			\end{subfigure}
			\begin{subfigure}{0.24\linewidth}
				\centering
				\includegraphics[width=\columnwidth]{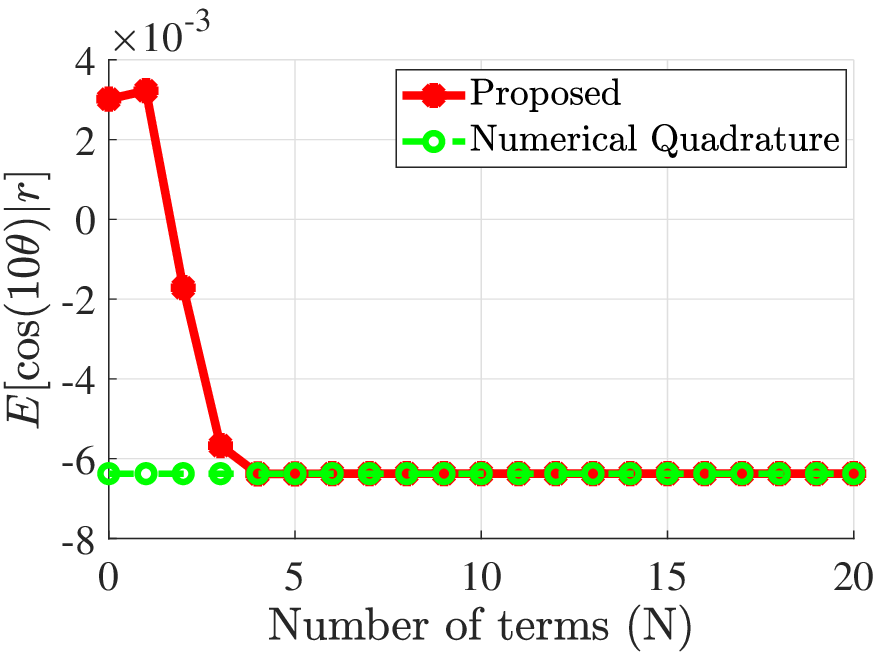}
				\caption{$m = 10$.}
			\end{subfigure}
			\caption{Computation of integral in Eqn. \eqref{eq_int_cosm} for various values of $m$.}
			\label{fig_cos_m}
		\end{figure*}	
		
		\begin{figure*}[t]
			\centering
			\begin{subfigure}{0.24\linewidth}
				\centering
				\includegraphics[width=\columnwidth]{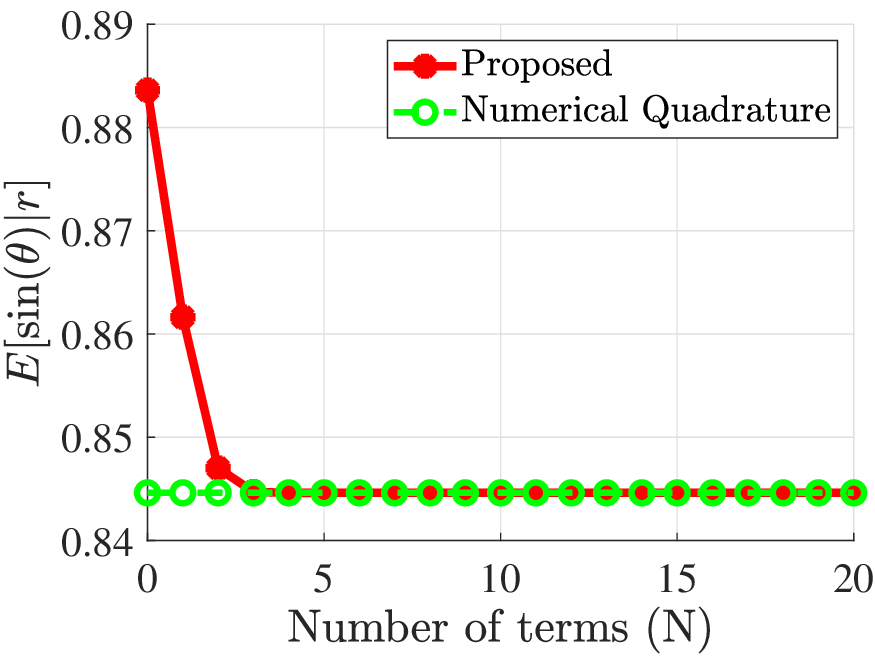}
				\caption{$m = 1$.}
			\end{subfigure}
			\begin{subfigure}{0.24\linewidth}
				\centering
				\includegraphics[width=\columnwidth]{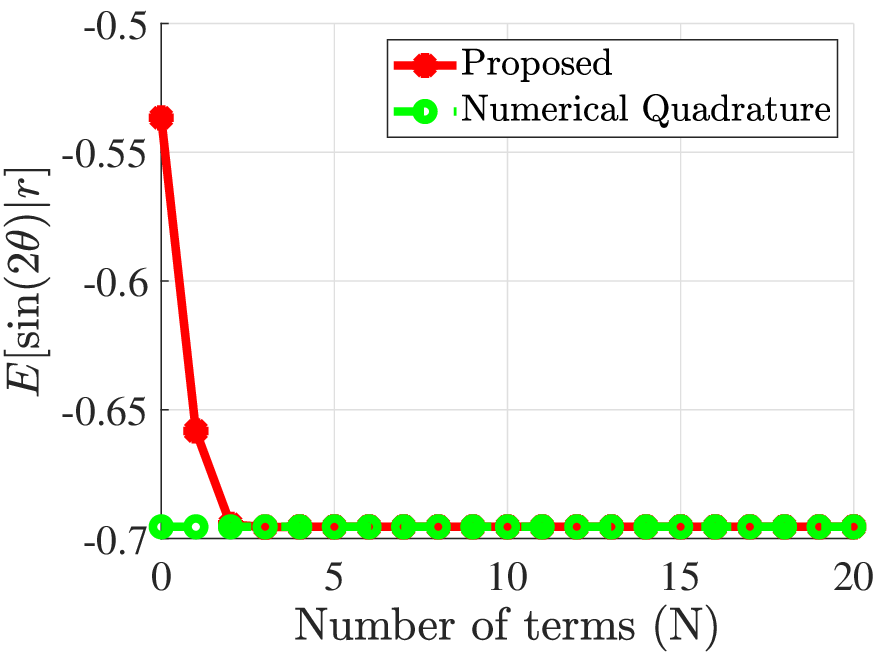}
				\caption{$m = 2$.}
			\end{subfigure}
			\begin{subfigure}{0.24\linewidth}
				\centering
				\includegraphics[width=\columnwidth]{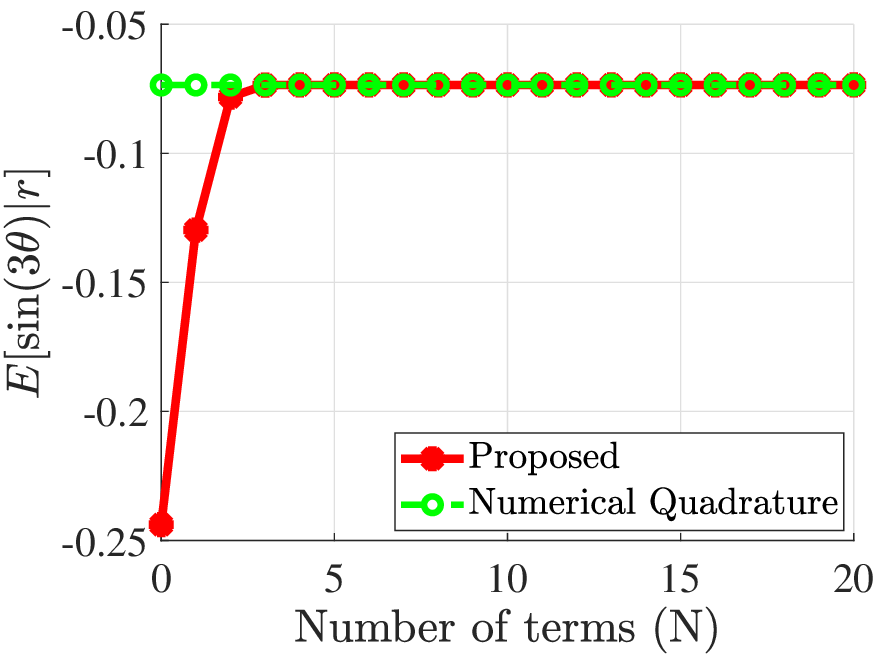}
				\caption{$m = 3$.}
			\end{subfigure}
			\begin{subfigure}{0.24\linewidth}
				\centering
				\includegraphics[width=\columnwidth]{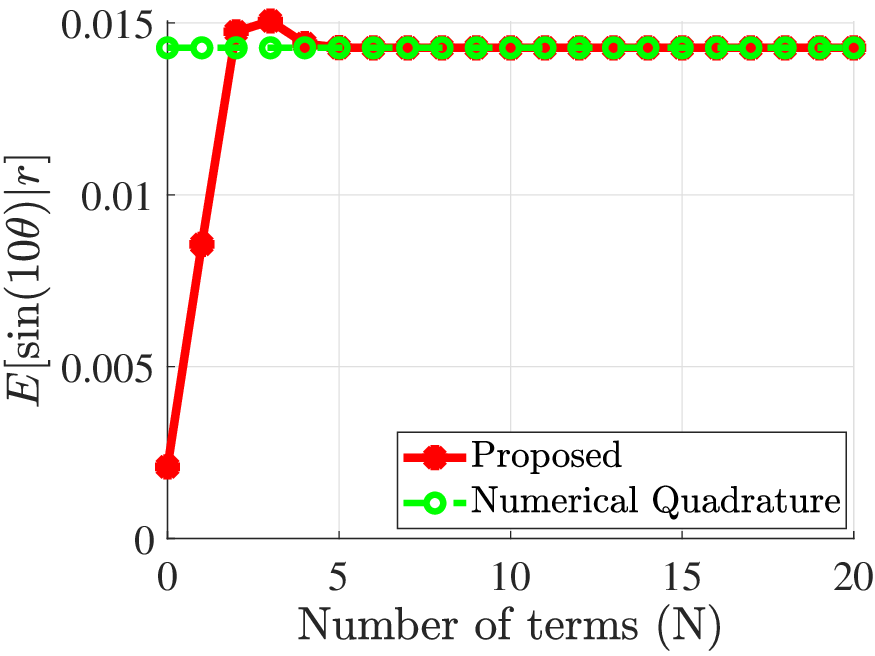}
				\caption{$m = 10$.}
			\end{subfigure}
			\caption{Computation of integral in Eqn. \eqref{eq_int_sinm} for various values of $m$.}
			\label{fig_sin_m}
		\end{figure*}	
		\section{Simulation Results} \label{sec_sim}
		The validity of the proposed formulation is simulated in two cases. First, to validate the accuracy of the solved integrals with a base-line method like quadrature. The other case presents the application of the proposed formulation to a real-life tracking scenario.
	
	\subsection{Comparison with Quadrature Rule}
	 	The integrals were evaluated with Matlab's\textsuperscript{\textcopyright} vectorized adaptive quadrature rule \cite{shampine2008vectorized} and found to converge for a few numbers of terms in summation. This simulation is based on the following values of assumed measurement mean $\hat{\mathbf{y}} = \begin{bmatrix}-11&20 \end{bmatrix}^T$ and measurement covariance $\mathbf{V} = \begin{bmatrix} 50 & -10 \\-10&50 \end{bmatrix}$.
	 	Range is calculated by adding zeros mean Gaussian noise with covariance $4\mathbf{I}_2$ to $\hat{\mathbf{y}}$. Integrals in eqns. \eqref{eq_int_cosm} and \eqref{eq_int_sinm} have been computed for a few values of $m$ using the proposed formulation in eqns. \eqref{eq_IntGen_nonsimp_cos} and \eqref{eq_IntGen_nonsimp_sin}. These values are plotted in Figs. \ref{fig_cos_m} and \ref{fig_sin_m} for the case $\cos$ and $\sin$ respectively against a finite number of terms $N$ employed in the summation of eqns. \eqref{eq_diagCos_pos} and \eqref{eq_diagSin_pos}.
	 	
\begin{figure}
		\centering
		\begin{subfigure}{0.48\linewidth}
			\centering
			\includegraphics[width=\linewidth]{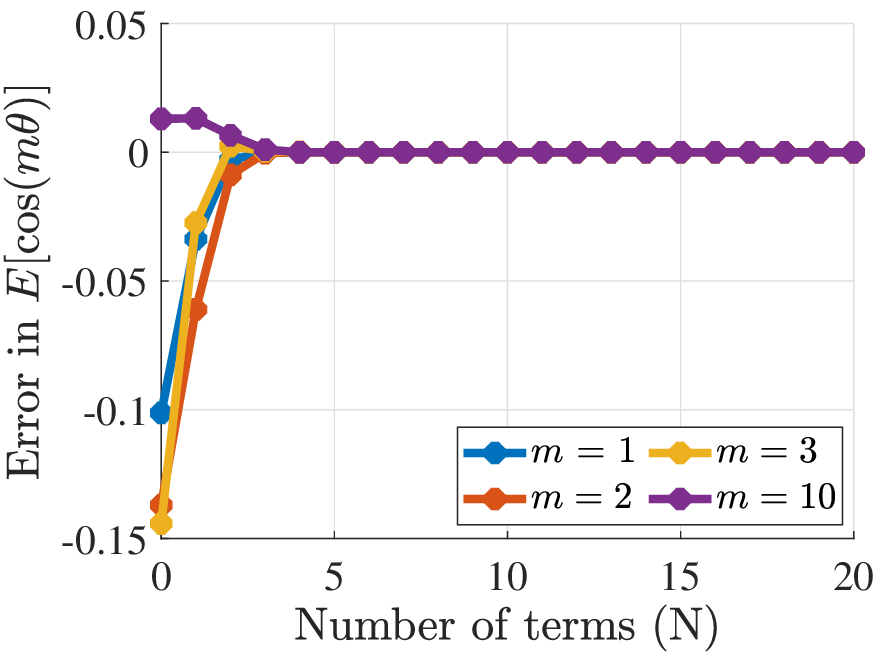}
			\caption{$\mathbb{E}[\cos(m\theta)|r]$}
		\end{subfigure}
		\begin{subfigure}{0.48\linewidth}
			\centering
			\includegraphics[width=\linewidth]{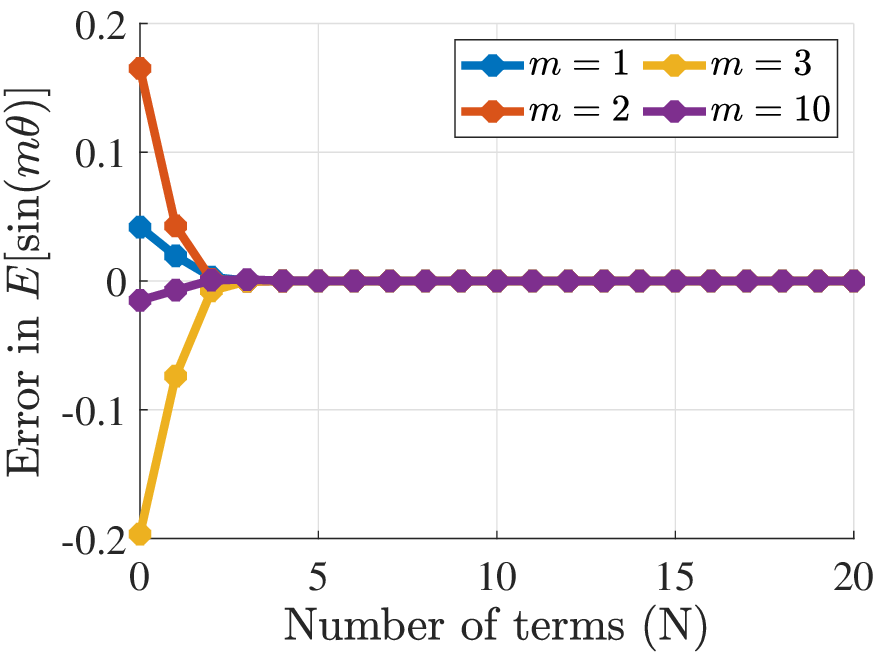}
			\caption{$\mathbb{E}[\sin(m\theta)|r]$}
		\end{subfigure}
		\caption{Error with respect to quadrature based computation.}
		\label{fig_error}
	\end{figure}

	 	The power of approximation is evident in the figures, where the proposed formulation converged with less than the first five terms of the infinite series solution. Since Matlab's quadrature implementation is adaptive, it doesn't take in number of steps as an argument, due to which it remains constant in Figs. \ref{fig_cos_m} and \ref{fig_sin_m}. The corresponding error is plotted in Fig. \ref{fig_error} and also tabulated in Table \ref{tab_error}. The table suggests that 10 terms are sufficient to calculate for significant accuracy. Beyond this, the error is incomprehensible for Matlab's\textsuperscript{\textcopyright} \texttt{double} precision values. The author recommends five terms are enough for most applications in target-tracking applications.
	 	
\begin{figure}
		\centering
		\begin{subfigure}{0.48\linewidth}
			\centering
			\includegraphics[width=\columnwidth]{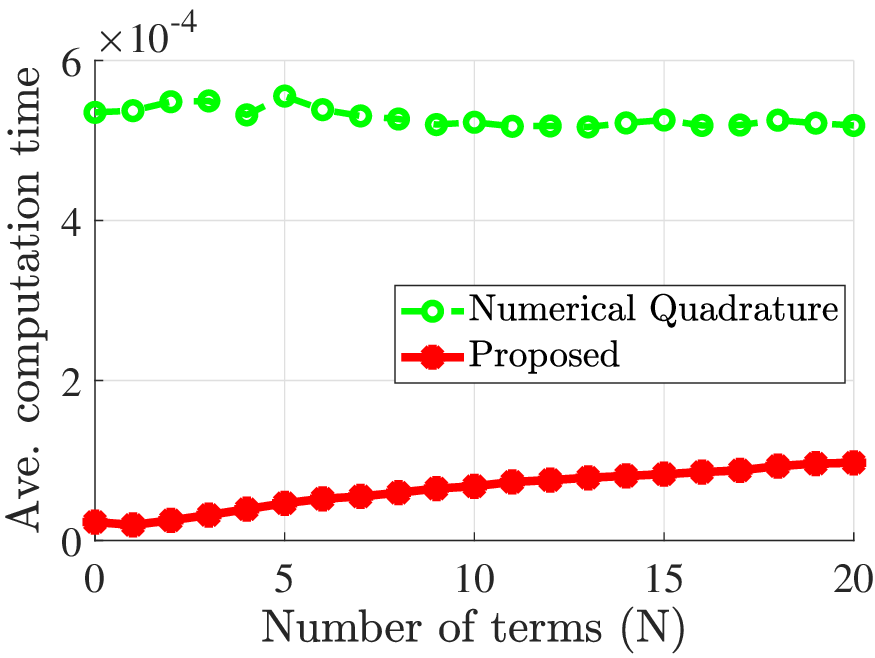}
			\caption{Without VPA}
			\label{fig_timeComp_noVPA}
		\end{subfigure}
		\begin{subfigure}{0.48\linewidth}
			\centering
			\includegraphics[width=\columnwidth]{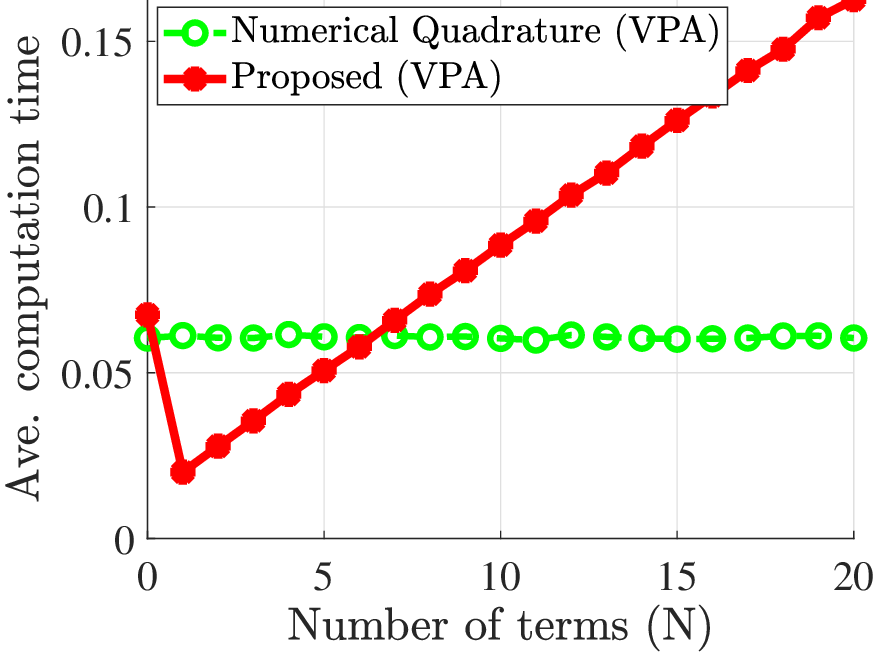}
			\caption{With VPA.}
			\label{fig_timeComp_VPA}
		\end{subfigure}
		\caption{Impact of Matlab's variable precision-arithmetic (VPA) on time-performance.}
		\label{fig_timePerf}
	\end{figure}

	 	The computation times for calculating $\mathbb{E}\left[\cos(\theta)\right]$ are plotted in Fig. \ref{fig_timePerf}. The advantage of using an exact solution is evident in Fig. \ref{fig_timeComp_noVPA}, where the quadrature rule performs 10 times slower as compared to the proposed formulation. For this example, Matlab's \texttt{double} precision was used, as the arguments were not too large. Since modified Bessel functions increase with argument, they can assume large values as $r$ increases (not shown here). This might return \texttt{Inf} or \texttt{NaN} while coding in Octave or similar systems. We employed Matlab's\textsuperscript{\textcopyright} variable-precision arithmetic (VPA) which provides the ability to apply mathematical operations to extremely large numbers without overflowing. This employs Matlab's symbolic toolbox\textsuperscript{\textcopyright} and hence slows down performance (default 32 bit precision used here). It can be seen that again, the time for quadrature rule is fairly constant as it doesn't depend on the number of terms as an input argument whereas the computation times in the case of exact formulation increases almost linearly. It can be observed from the plot that the proposed method is still advantageous to use below five terms of summation, beyond which the quadrature can be used. 
		 
		 \begin{table*}
		 	\centering
		 	\caption{Error in computation (m = 1).}
		 	\label{tab_error}
		 	\begin{tabular}{ccccccc}
		 		\toprule
		 		& N = 0	& 2	& 5 & 10 & 15 & 20 \\
		 			\midrule
			 Error in ${\mathbb{E}[\cos(\theta)|r]}$ & -0.07 & $-2.9\times10^{-3}$ & $-1.05\times10^{-8}$ & $7.63\times10^{-15}$ & $-3.12\times10^{-17}$ & $-3.12\times10^{-17}$\\
			 Error in ${\mathbb{E}[\sin(\theta)|r]}$ & 0.04 & $3.1\times10^{-3}$ & $1.74\times10^{-7}$ & $1.46\times10^{-15}$ &  $1.66\times10^{-17}$ & $1.66\times10^{-17}$\\
		 		\bottomrule
		 	\end{tabular}
		 \end{table*}

	\subsection{Single-Sensor Range-Only Target Tracking Scenario}
	A single-sensor range-only tracking scenario is non-trivial to construct due to challenges in observability. We choose an existing scenario which is discussed in \cite{pillon2016observability} as conditionally observable. The observability is such that it can present a ghost target which can cause divergence in estimation, but this effect can be nullified by taking a single azimuth measurement of appropriate standard deviation at the beginning, which also constructs the prior density. Such an approach is also discussed in \cite{ristic2002target} where the RADAR begins in a scan mode and subsequently switches to inverse synthetic aperture radar (ISAR) mode to retrieve high resolution range-only measurements. In this simulation as well, we capture a single azimuth measurement to construct the prior.
	
	The total simulation time is 30 minutes, with the rest of the parameters presented in Table \ref{tab_simParam}. The target is assumed to be traveling with a nearly constant velocity model injected with a high process noise. Initially, it is situated 10 kilometers away from the sensor and starts moving towards it with a heading of $225^\circ$. The target speed is taken to be $15$ knots. The observer, initially located at $(0,0)$ kilometers, starts moving with a heading direction of $170^\circ$ for $15$ minutes and abruptly maneuvers by changing its heading to $304^\circ$. Throughout the traversal, the observer speed is fixed at $5$ knots. Note that all angles are measured counter-clockwise from the local x-axis (or easting in local ENU coordinates).
	
	The relative target motion at time $k$ is modeled using the following equation,
	\begin{align}
		\mathbf{x}^r_k = \mathbf{F}_k\mathbf{x}^r_{k-1} + \mathbf{F}_k\mathbf{x}^o_{k-1} - \mathbf{x}^o_{k} + \mathbf{w}_k
	\end{align}
	where $\mathbf{x}^r_k $ is the target state relative to the observer state $\mathbf{x}^o_{k}$. $\mathbf{F}_k$ is the linear state transition matrix, and $\mathbf{w}_k$ is the zero-mean process noise with covariance $\mathbf{Q}_k$,	
	\begin{align}
		\mathbf{F}_k = \begin{bmatrix}
			\mathbf{I}_2 &  T\mathbf{I}_2 \\
			\mathbf{0}_2 & \mathbf{I}_2
		\end{bmatrix}, \quad \mathbf{Q}_k &= \tilde{q}\begin{bmatrix}
			\frac{ T^3}{3} \mathbf{I}_2 & \frac{ T^2}{2} \mathbf{I}_2 \\
			\frac{ T^2}{2} \mathbf{I}_2 &  T \mathbf{I}_2
		\end{bmatrix}
	\end{align}
	where $\tilde{q}$ is the process noise intensity and $T$ is the sampling-time. For measurements, the following conventional range-only measurement (noise-after-norm) is employed to check the versatility of proposed algorithm to existing systems,
	\begin{align}
		r_k = \sqrt{x_k^2 + y_k^2}+ v_k
	\end{align}  
	where $v_k$ is a zero-mean measurement noise with covariance $\sigma_r^2$. 
	
\begin{figure}
			\centering
				\includegraphics[width=0.85\columnwidth]{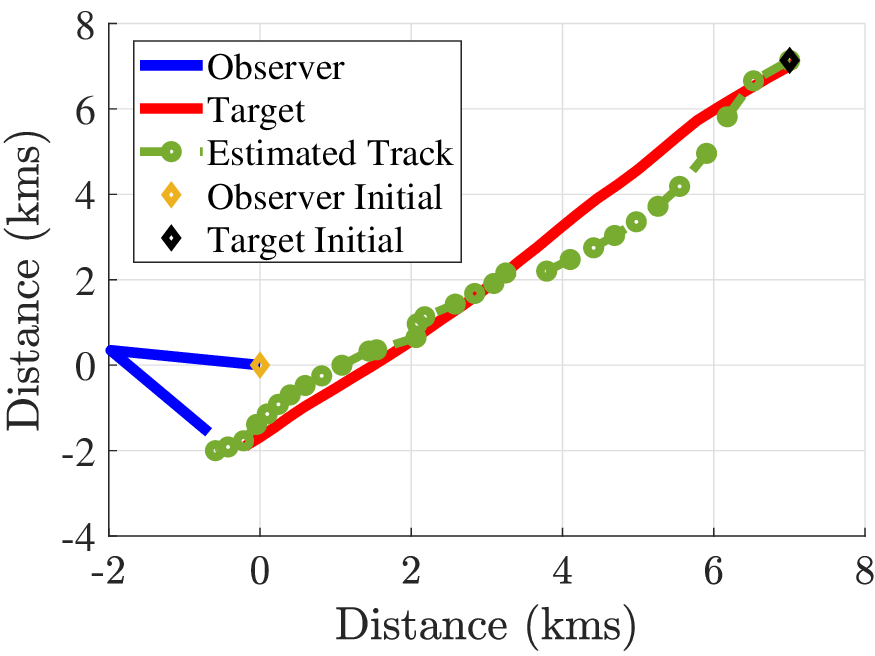}
				\caption{True and estimated trajectories.}
				\label{fig_scene}
		\end{figure}

	The scenario is presented in Fig. \ref{fig_scene} along with the estimated track using the proposed approach. Ideally, without process noise, the observer lies along the target path, but the high process noise causes deviations in the target trajectory. As shown, the proposed approach faithfully tracks the target, with convergence achieved after observer maneuver.
	
\begin{figure}
			\centering
			\includegraphics[width=0.85\columnwidth]{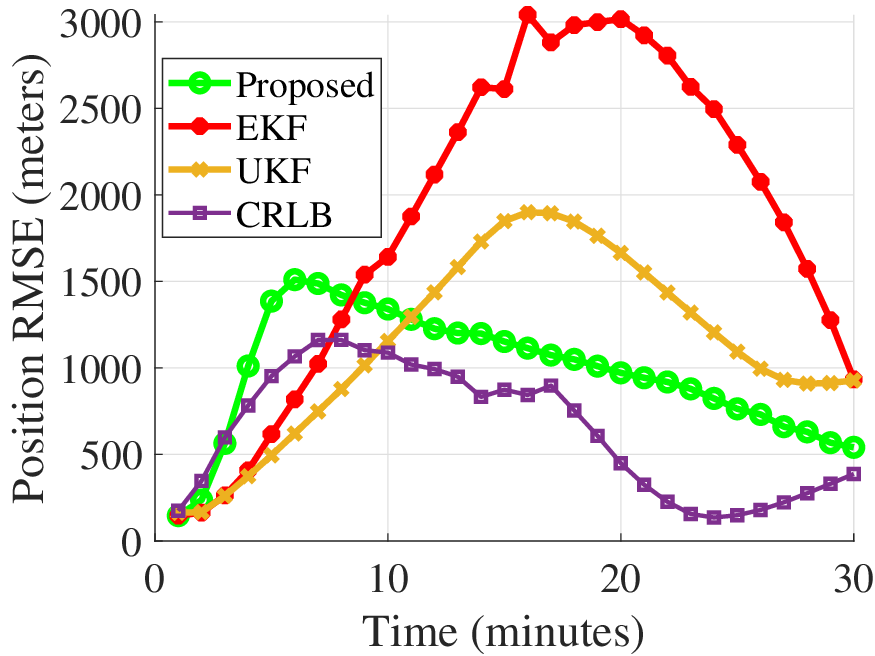}
			\caption{Position RMSE.}
			\label{fig_pos_rmse}
		\end{figure}

	For performance evaluation, we used root-mean-square error (RMSE) for position and velocity, as well as the normalized estimation-error squared (NEES), averaged over 100 Monte-Carlo runs. The RMSE position is shown in Fig. \ref{fig_pos_rmse}, where the proposed approach stands out in comparison to the extended Kalman filter (EKF) and the unscented Kalman filter (UKF) \cite{wan2001unscented}. Cr\'amer-Rao lower bound (CRLB) derived from \cite{tichavsky1998posterior} is also presented as a benchmark.
	
	Both UKF and EKF use a linear approach by approximating the conditional distribution as Gaussian and applying it to a linear Kalman filter. Since the range standard-deviation is low, the uncertainty region cannot be sufficiently approximated as an ellipsoid, due to which the minimum mean square error (MMSE) estimate is no longer captured by the mean of the Gaussian approximated posterior. 
	
\begin{figure}
			\centering
			\includegraphics[width=0.85\columnwidth]{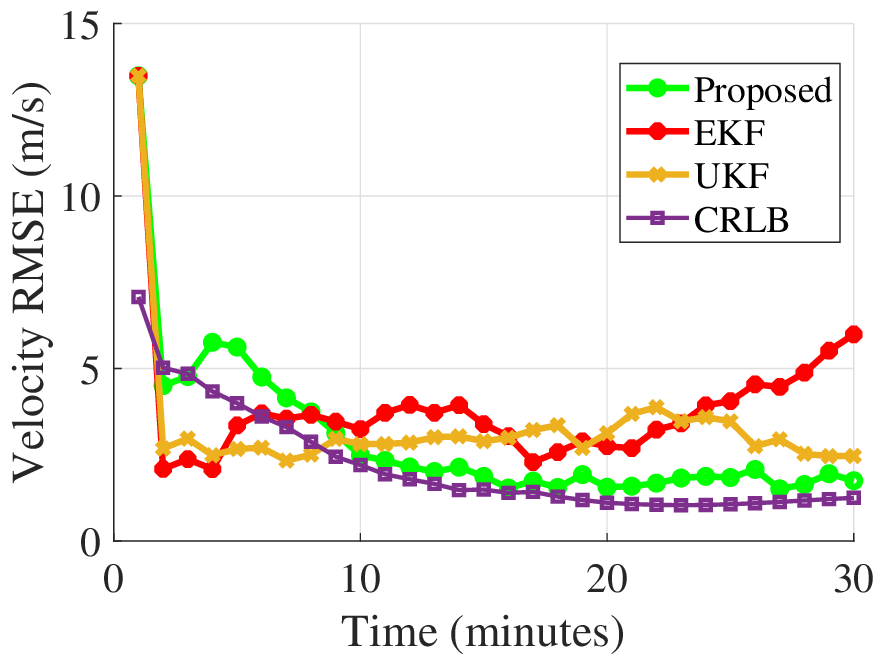}
			\caption{Velocity RMSE.}
			\label{fig_vel_rmse}
		\end{figure}

	A similar trend can be seen in the velocity RMSE plot in Fig. \ref{fig_vel_rmse}. The proposed algorithm stands out even when Gaussian approximation is used in place of mixture posterior density, and almost reaches CRLB. Note that unlike the position RMSE plot, the CRLB does not start from the same initial point, which is due to the fact that single-point initialization was used for estimating the prior. 
	
\begin{figure}
			\centering
			\includegraphics[width=0.85\columnwidth]{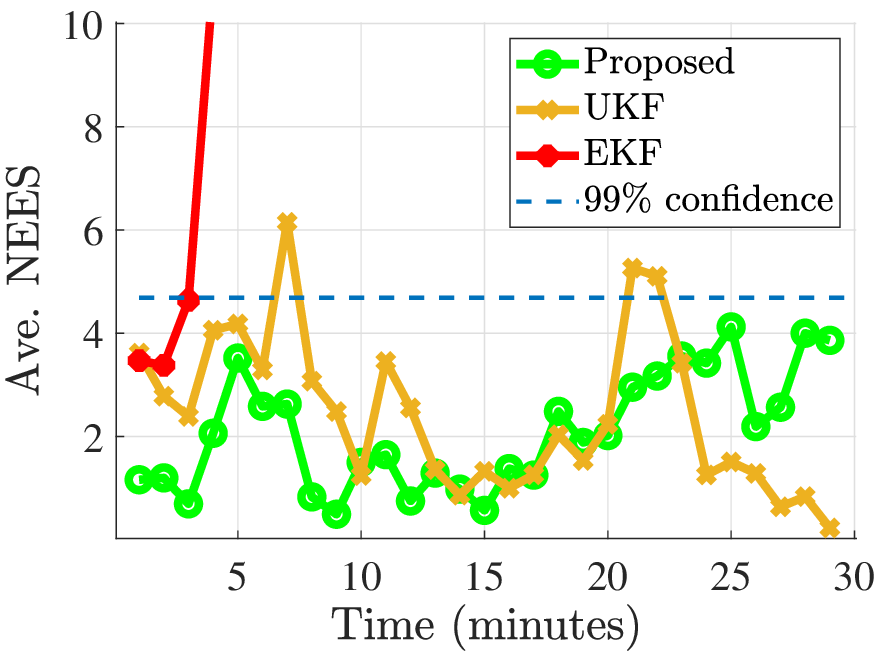}
			\caption{Average NEES.}
			\label{fig_ave_nees}
		\end{figure}

	The last component of the simulation result is the NEES plot in Fig. \ref{fig_ave_nees} where the EKF is shown to be completely inconsistent in comparison to UKF and the proposed approach which are rather consistent due to Gaussian approximation of the posterior.
	
	\begin{table}[t]
		\centering
		\caption{Simulation Parameters}
		\label{tab_simParam}
		\begin{tabular}{ll}
			\toprule
			\textbf{Parameters} & \textbf{Value} \\
			\midrule
			Sampling time & 60 seconds. \\
			Simulation time & 30 minutes. \\
			Initial target pos. & $\left[7072.1,7072.1\right]$ meters. \\
			Target heading & $225^\circ$. \\
			Init. observer heading & $170^\circ$.\\
			Final observer heading & $304^\circ$. \\
			Target speed & 15 knots. \\
			Observer speed & 5 knots. \\
			$\tilde{q}$ & $10^{-3}$ $\text{m}^2/\text{sec}^3$.\\
			$\sigma_r$ & 10 meters. \\
			$\sigma_\theta$ & $1^\circ$. \\
			No. of terms in summation $(N)$ & 5 \\
			Highest circular moment $(M)$ & 10\\
			Components in Dirac approximation $(L)$ &  8\\
			\bottomrule
		\end{tabular}
	\end{table}  
		
		\section{Conclusion} \label{sec_concl}
		In this paper, we focused on a specific problem in the domain of 2D range-only tracking (ROT). We show that the Fourier type integral of the azimuth conditional density exists as an infinite sum, which can produce solutions for a few numbers of summing elements. We rigorously derive the generalized density of range and then use the solution to derive the expectation of the form $\mathbb{E}\left[\cos(m\theta)\right]$ and $\mathbb{E}\left[\sin(m\theta)\right]$ for an arbitrary positive integer $m$. Using such moments, it is possible to accurately find deterministic samples of the conditional azimuth density. The formulation was compared using numerical quadrature, which proved that the proposed approach converges swiftly at a fraction of cost.
		
		The rest of the filtering algorithm follows from the standard total probability theorem, wherein the discrete azimuth distribution was employed, resulting in a Gaussian mixture distribution as the posterior. The results provide evidence of the benefit of using exact circular moments of the conditional density, thus, capturing the multi-modality of state uncertainty. 
		
		As part of the future research, the authors would examine better approaches for calculating Bessel function ratios, which even though existing in the current literature, are almost equivalent in terms of computation time, to using MATLAB's\textsuperscript{\textcopyright} variable precision arithmetics. Also, the optimization based approach of calculating sigma-points can be refined. Both of these approaches, when employed, would result in a much robust and faster filtering method for 2D range-only target tracking problems.
		
		\appendix
		\subsection{Proof of Eqn. \eqref{eq_rangeDensity_integral}}
		It has been proved in \cite{weil1954distribution}, that the following integral can be computed in closed form,
		\begin{align}
		 &\int_0^{2\pi} \exp\bigl[r_k \norm{p,q} \cos(\theta_k-\phi_1) +
		 A_3r_k^2 \cos(2\theta_k)\bigr] d\theta_k \notag \\
		 &\quad = 2\pi \bigl[ I_0\left(A_3r_k^2\right)I_0\left(r_k\sqrt{A_1^2 + A_2^2}\right) + \notag \\
		 &\quad 2\sum_{k=1}^{\infty} I_k\left(A_3r_k^2\right)I_{2k}\left(r_k\sqrt{A_1^2 + A_2^2}\right)\cos(2k\psi) \bigr]. \label{eq_diagRange_integral}
		\end{align}
		The proof of Eqn. \eqref{eq_rangeDensity_integral} follows from the fact that the integral can be converted to the diagonal case in the L.H.S of equation \eqref{eq_diagRange_integral} by the substituting $2\theta_k + \phi_2 = 2u$. After converting to the standard form, the proof is trivial. 
		
		\subsection{Proof of Eqns. \eqref{eq_diagCos} and \eqref{eq_diagSin}}
		We wish to compute the integral $\text{I}_\text{diag,cos}(m)$,
		\begin{align}
			\int_0^{2\pi}\exp\bigl[r_k \norm{p,q} \cos(\theta_k-\phi_1) &+ A_3r_k^2 \cos(2\theta_k)\bigr]  \notag \\
			&\times\cos(m\theta_k)d\theta_k,
		\end{align} 
 		which can be rewritten in the standard form as,
 		\begin{align}
 			\int_0^{2\pi}\exp\bigl[A_1r_k\cos(\theta_k) + A_2r_k\sin(\theta_k) &+ A_3r_k^2 \cos(2\theta_k)\bigr] \notag \\ &\times\cos(m\theta_k)d\theta_k,
 		\end{align}
 	where $A_1$ and $A_2$ are given in Eqn. \eqref{eq_const_A1_A2}. Substituting the following identities for the Bessel function in the equation above,
 	\begin{align}
 		\exp\left[z\sin(\theta)\right]&= \sum_{n = -\infty}^\infty I_n(z)\exp\left[ni\left(\theta + \frac{3\pi}{2}\right)\right], \\
 		\exp\left[z\cos(\theta)\right] &= \sum_{n = -\infty}^\infty I_n(z)\exp\left[ni\theta\right],
 	\end{align}
 	where $I_n(z)$ is the modified Bessel function of the first kind and order $n$. The result is (removing the subscript $k$ for time),
 	\begin{align}
 		\text{I}_\text{diag,cos}(m) = \int_{0}^{2\pi}\cos(m\theta)d\theta\sum_{j=-\infty}^\infty\sum_{k=-\infty}^\infty\sum_{l=-\infty}^\infty I_j(A_3r^2) \notag \\
 				\times I_k(A_1r)I_l(A_2r)\exp\left[i\theta_k\underbrace{(2j+k+l)}_{M} + i\underbrace{\frac{3\pi l}{2}}_{L} \right], \label{eq_append_M_L}
 	\end{align}
 	Moving the integral inside the summation, and using Euler's identity for $\cos(m\theta)$,
 	\begin{align}
 	&\text{I}_\text{diag,cos}(m) = \sum_j\sum_k\sum_l\frac{1}{2}I_j(A_3r^2)I_k(A_1r)I_l(A_2r) \quad \times \notag  \\
 							&\int_0^{2\pi} \biggl[ \exp(i\theta(M+m) + iL) + \exp(i\theta_k(M-m) + iL)\biggr]d\theta, \label{eq_cosDiag_EulerStep}
 	\end{align} 
 	where $M$ and $L$ are indicated in Eqn. \eqref{eq_append_M_L}. Taking the real part of the above equation and using the following identities for an integer $m$,
 	\begin{subequations}\label{eq_trig_identities}
 	\begin{align}
 		\int_0^{2\pi} \cos(m\theta) &= \begin{cases}
 											2\pi \quad &\text{m = 0} \\
 											0 \quad &\text{otherwise}
 										\end{cases} \\
 		\int_0^{2\pi} \sin(m\theta) &= 0,					
 	\end{align}
 	\end{subequations}
 	which means that the integral is non-zero only for following cases,
 	\begin{align}
 		(M \pm m) = 0,\quad l = 2n,
  	\end{align}
 	The resulting integral is then,
 	\begin{align}
 		\text{I}_\text{diag,cos}(m) &= \frac{2\pi}{2}\sum_jI_j(A_3r^2)\quad\times \notag \\
 		&\quad \sum_{n=\infty}^\infty (-1)^n\bigr[ I_{2n+2j+m}(A_1r)I_{2n}(A_2r)\notag \\
 									&\quad + I_{2n+2j-m}(A_1r)I_{2n}(A_2r) \bigl], \label{eq_b4AddTheor_cos}
 	\end{align}
 	Now, use of the following addition theorems \cite{watson1922treatise} is made, which reduces the above expression to a single sum. 

 	\begin{align}
 		\sum_{q=-\infty}^\infty (-1)^q I_{p+2q}(Z)I_{2q}(z) &= I_p\left(\sqrt{Z^2 + z^2}\right)\cos(p\psi),\label{eq_addTheorem_cos}
 	\end{align}
 	where $\tan(\psi) = z/Z$. Eqn. \eqref{eq_diagIntegral_cos} follows from the substitution of Eqn. \eqref{eq_addTheorem_cos} in \eqref{eq_b4AddTheor_cos}. 
 	
 	The proof for $\text{I}_\text{diag,sin}(m)$ follows in a similar manner except that Euler identity for $\sin$ is used in Eqn. \eqref{eq_cosDiag_EulerStep},
 	\begin{align}
 		\text{I}_\text{diag,sin}(m) &= \sum_j\sum_k\sum_l\frac{1}{2}I_j(A_3r^2)I_k(A_1r)I_l(A_2r) \quad \times \notag  \\
 		&\int_0^{2\pi} \biggl[ \exp(i\theta(M+m) + i(L-\frac{\pi}{2})) \notag \\
 		&- \exp(i\theta_k(M-m) + i(L-\frac{\pi}{2}))\biggr]d\theta. \label{eq_sinDiag_EulerStep}
 	\end{align}
 	The integral is non-zero for the following cases,
 	\begin{align}
 		(M \pm m) = 0,\quad l = 2n+1.
 	\end{align}
 	Making use of the identities in Eqn. \eqref{eq_trig_identities}, the integral reduces to,
 	\begin{align}
 		\text{I}_\text{diag,sin}(m) &= \frac{2\pi}{2}\sum_jI_j(A_3r^2)\quad\times \notag \\
 		&\quad \sum_{n=\infty}^\infty (-1)^{n+1}\bigl[ I_{2n+2j+m+1}(A_1r)I_{2n+1}(A_2r)\notag \\
 		&\quad - I_{2n+2j-m+1}(A_1r)I_{2n+1}(A_2r) \bigr]. \label{eq_b4AddTheor_sin}
 	\end{align}
 	Again, we make use of the following addition theorem which reduces the expression to a single summation,
 	\begin{align}
 		\sum_{q=-\infty}^\infty (-1)^{q+1} I_{p+2q+1}(Z)I_{2q+1}(z) &= I_p\left(\sqrt{Z^2 + z^2}\right)\sin(p\psi)\label{eq_addTheorem_sin}.
 	\end{align}
 	Equation \eqref{eq_diagIntegral_sin} is the result of substitution of Eqn. \eqref{eq_addTheorem_sin} in Eqn. \eqref{eq_b4AddTheor_sin}.
	 	 
	 \bibliography{MAIN_paper_3.bib}
	 \bibliographystyle{ieeetr}

\end{document}